\documentclass[reprint,superscriptaddress,preprintnumbers,amsmath,amssymb,aps,prd,tightenlines,longbibliography,nofootinbib]{revtex4-2}
\usepackage{graphicx} 
\usepackage{hyperref}
\usepackage[utf8]{inputenc}
\usepackage[LGR,T1]{fontenc}
\usepackage{textgreek}
\usepackage{amsbsy}

\renewcommand{\v}[1]{\boldsymbol{#1}}
\newcommand{\lh}{\mathcal{L}}
\newcommand{\Nmax}{n_{\mathrm{m}}}
\renewcommand{\d}{\mathrm{d}}
\newcommand{\binomp}{\mathcal{B}}
\newcommand{\CDF}{\mathrm{CDF}}

\begin{document}

\title{Kilo-scale point-source inference using Parametric Cataloging}
\begin{abstract}
The estimation of the number of point-sources in the sky is one the oldest problems in astronomy, yet an easy and efficient method for estimating the uncertainty on these counts is still an open problem.
Probabilistic cataloging solves the general point-source inference problem, but the trans-dimensional nature of the inference method requires a bespoke approach that is difficult to scale.
Here it is shown that probabilistic cataloging can be performed in a fixed-dimensional framework called Parametric Cataloging under mild assumptions on some of the priors.
The method requires only a simple reparameterization of the flux coordinates, yielding an accessible method that can be implemented in most probabilistic programming environments. As the parameter space is fixed-dimensional, off the shelf gradient based samplers can be employed which allows the method to scale to tens of thousands of sources.
\end{abstract}
 

\author{Gabriel H. Collin}
\email{gabriel.collin@adelaide.edu.au}
\affiliation{Department of Physics, The University of Adelaide, Adelaide, SA 5005, Australia}

\maketitle

\section{Introduction}

Cataloging is the process of turning data -- such as astronomical images or lists of events -- into a set of point sources that generate them.
The location, flux and other parameters of each source in this set are estimated; and, when the sources are bright and separated we need only quantify the uncertainty on those parameters.
There exists a regime when sources are dim or crowded, where a single set of source parameters and their uncertainties is not enough to quantify the total uncertainty of the problem.
In this regime, it is difficult to distinguish whether a putative source exists or is mimicked by a statistical fluctuation in the data --- the number of sources has significant uncertainty.

Probabilistic cataloging \cite{Hogg:2010ip,Brewer:2012gt,Daylan:2016tia} is a Bayesian approach to quantifying uncertainty in both the number of sources and the parameters of those sources.
The likelihood, $\lh(\v{d}|n,\{\v{\phi}_i\})$, for the observed data, $\v{d}$, is conditioned on the number of sources, $n$, and a set of $n$ vectors, $\{\v{\phi}_i\}$, where each $\v{\phi}_i$ is the vector of parameters for a single source. 
The goal is to draw samples from the posterior
\begin{equation}
    p(n,\{\v{\phi}_i\},\v{\theta}|\v{d}) = \frac{ \lh(\v{d}|n,\{\v{\phi}_i\}) \pi(\v{\theta}) \pi(n|\v{\theta}) \left( \prod_{i=1}^n \pi(\v{\phi}_i | \v{\theta}) \right) }{\lh(\v{d})},
\end{equation}
where $\v{\theta}$ is a vector of hyperparameters -- such as mean number of sources or average flux per source -- with prior $\pi(\v{\theta})$; $\pi(n|\v{\theta})$ is the prior on the number of sources; $\pi(\v{\phi}|\v{\theta})$ is the prior on the parameters of one source (assumed identical across sources); and $\lh(\v{d})$ is the marginal likelihood.
Each sample drawn from this posterior is a triple $(n, \{\v{\phi}_i\}, \v{\theta})$ that represents a single catalog.
The ensemble of posterior samples therefore represents equally probable catalogs that could be constructed from the data.

Note that these samples are not fixed-dimensional; the size of the set $\{\v{\phi}_i\}$ varies with $n$.
This precludes the naive application of Markov Chain Monte Carlo (MCMC) to sample from the posterior, as most MCMC algorithms are designed only to sample from fixed-dimensional spaces.
An exception to this is Reversible Jump Markov Chain Monte Carlo (RJMCMC) \cite{Green:1995mxx,richardsonBayesianAnalysisMixtures1997a}, a transdimensional MCMC that can adjust the dimensionality of the space from which it samples.

RJMCMC forms the backbone of current applications of probabilistic cataloging \cite{Brewer:2012gt,Brewer:2015yya,jonesDISENTANGLINGOVERLAPPINGASTRONOMICAL2015,brewerFastBayesianInference2015,Daylan:2016tia,portilloImprovedPointsourceDetection2017,Daylan:2017kfh,sottosantiDiscoveringLocatingHighEnergy2019,federMultibandProbabilisticCataloging2020,Butler:2021val,costantinBayesianMixtureModelling2022,federPCATDEReconstructingPointlike2023}, but has two major limitations.
Firstly, the transdimensional updates require carefully hand-crafted proposal functions to ensure efficient proposals.
This bespoke complexity is a barrier to entry for new applications of probabilistic cataloging.
Secondly, RJMCMC does not scale to large numbers of parameters. 
Although this problem can be partially ameliorated for updates to the source parameters \cite{Daylan:2016tia}, updates to hyperparameters or diffuse components still pose a scaling difficulty.
The nature of trans-dimensional updates precludes the use of gradient-based sampling strategies -- such as Hamiltonian Monte Carlo (HMC) \cite{Duane:1987de,Neal:2011mrf,Betancourt:2017ebh} -- that scale to high dimensions.

In this paper, it is shown that probabilistic cataloging can be performed in a fixed-dimensional framework called Parametric Cataloging.
This requires a binomial assumption for the prior on $n$, along with some mild regularity assumptions on the flux prior.
This addresses both concerns with using RJMCMC.
Firstly, parametric cataloging requires only a simple reparameterization of the flux coordinates, yielding an accessible method that can be implemented in most probabilistic programming environments.
Secondly, as the parameter space is fixed-dimensional, off the shelf gradient based samplers can be employed which allows the method to scale to tens of thousands of sources.

Other approaches to estimating the uncertainty on the number of sources exist, such as modeling the problem using a Dirichlet process mixture \cite{Sottosanti:2021eid}.
When probabilistic catalogs are not of interest, the population hyperparameters can be modeled directly through likelihood \cite{Collin:2021ufc,Lee:2015fea,Malyshev:2011zi,Lee:2008fm} or machine-learning based approaches 
\cite{Caron:2017udl,List:2020mzd,List:2021aer,Mishra-Sharma:2021oxe,Caron:2022akb,Butter:2023piw,Amerio:2023uet,Christy:2024gsl,Wolf:2024oqb,List:2025qbx}.

Point source inference can be considered to be a case of the more general problem of ``clustering'' in the statistical literature \cite{gormleyModelBasedClustering2023}, specifically a mixture-model.
There are many approaches to this problem \cite{celeuxComputationalSolutionsBayesian2018}; to this work, the closest of which is overfitted mixture-models \cite{rousseauAsymptoticBehaviourPosterior2011,malsiner-walliModelbasedClusteringBased2016}.
These fit a large number of components (sources) to the data, and use a specific choice of shrinkage prior on the component weights (fluxes) to approximately empty the redundant components.
A Dirichlet prior is chosen to heavily prefer low weights, thus driving the model to have as many near-zero weights as possible.
This work, in comparison, also uses a large number of components but uses a reparameterization to implement a spike-and-slab \cite{mitchellBayesianVariableSelection1988} type prior on the weights that allows a more general specification of flux prior and exact zeroing-out of redundant components.
Unlike the standard spike-and-slab approach, which uses discrete indicator variables that are computationally demanding, parametric cataloging is implemented in a purely continuous parameter space, utilizing a reparameterization trick to generate the spike-and-slab.

While inspired by Nonparametric Hamiltonian Monte-Carlo (NP-HMC) \cite{pmlr-v139-mak21a}, this approach is dramatically different in implementation.
NP-HMC solves general non-parametric inference problems using a novel and specialized variant of HMC.
In contrast, parametric cataloging solves a specialized inference problem by casting the problem in a fixed-dimensional (parametric) framework that permits the use of any sampler.

\section{Parametric cataloging}

We start with the mixture-of-likelihoods formulation of probabilistic cataloging where the mean number of sources $N$ and flux parameters for each source $F_i$ have been separated out:
\begin{multline}
    \lh(\v{d}|N,\v{\theta})
        = \sum_{n=0}^{\infty} \pi(n|N) \\ \left(\prod_{i=1}^n  \int \d\v{\phi}_i \pi(\v{\phi}_i|\v{\theta}) \int \d F_i \pi(F_i|\v{\theta})\right) 
        \lh(\v{d}|n, \{\v{\phi}_i\}, \{F_i\}). \label{eq:mixture_pc}
\end{multline}
In this representation, the integrals over the $\v{\phi}_i$ vectors are performed through Monte Carlo sampling --- hence forming probabilistic catalogs.

We now let the prior over $n$ be given by a binomial distribution $\binomp$:
\begin{equation}
    \pi(n|N) = \begin{cases} \binomp(n|\Nmax,N/\Nmax) & n \leq \Nmax \\ 0 & \text{otherwise} \end{cases}
\end{equation}
where $\Nmax$ is the maximum number of sources under consideration and
\begin{equation}
    \binomp(n|\Nmax,N/\Nmax) = {\Nmax \choose n} \left(\frac{N}{\Nmax}\right)^n \left(1-\frac{N}{\Nmax}\right)^{\Nmax - n}.
\end{equation}
This limits the sum in \eqref{eq:mixture_pc} to $\Nmax$ terms.

\begin{figure}
  \centering
  \includegraphics[page=1,width=0.8\columnwidth]{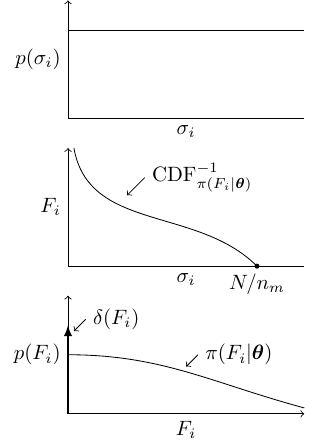}
  \caption{Above: A uniform prior is required for the new $\sigma_i$ parameters. Middle: The $\sigma_i$ parameters are transformed by $f_{N,\v{\theta}}$ into flux parameters $F_i$. Below: The transformed distribution has a spike component at the origin denoted by the bold arrow, and a slab component that follows the flux prior.}
  \label{fig:sspriorregular}
\end{figure}

A crucial step is in recognizing that the likelihood can be padded with additional sources up to $\Nmax$ as long as those sources have zero flux:
\begin{multline}
    \lh(\v{d} | n, \{\v{\phi}_i\}, \{F_i\}) = \lh(\v{d} | \Nmax, \{\v{\phi}_1,\ldots,\v{\phi}_n,\v{\phi}_{n+1},\ldots\}, \\ \{F_1,\ldots,F_n,0,\ldots\}),
\end{multline}
where the additional source parameters $\v{\phi}_{n+1},\ldots$ are arbitrary.
With this we can pull the likelihood out of the sum in \eqref{eq:mixture_pc} with the help of some Dirac delta distributions:
\begin{multline}
    \lh(\v{d}|N,\v{\theta}) = \\ \left(\prod_{i=1}^{\Nmax}  \int \d\v{\phi}_i \pi(\v{\phi}_i|\v{\theta}) \int \d F_i \right) \lh(\v{d}|\Nmax,\{\v{\phi}_i\},\{F_i\})  \\
    \sum_{n=0}^{\Nmax} \binomp(n|\Nmax,N/\Nmax) \left(\prod_{i=1}^{n} \pi(F_i|\v{\theta}) \right) \left( \prod_{i=n+1}^{\Nmax} \delta(F_i) \right). \label{eq:binom_expanded}
\end{multline}

The other critical property of the likelihood is that it is invariant under permutation of the source parameters.
We can swap any $(\v{\phi}_i,F_i)$ with any $(\v{\phi}_j,F_j)$ and the likelihood retains the same value.
With this property, we can use the binomial theorem to factorize the sum in \eqref{eq:binom_expanded}:
\begin{multline}
    \lh(\v{d}|N,\v{\theta}) = \\ \left(\prod_{i=1}^{\Nmax}  \int \d\v{\phi}_i \pi(\v{\phi}_i|\v{\theta}) \int \d F_i \right) \lh(\v{d}|\Nmax,\{\v{\phi}_i\},\{F_i\})  \\
    \prod_{i=1}^{\Nmax} \left( \frac{N}{\Nmax} \pi(F_i|\v{\theta}) + \left(1-\frac{N}{\Nmax}\right)\delta(F_i) \right). \label{eq:binom_factored}
\end{multline}

Thus it is revealed that probabilistic cataloging with this particular choice of prior on $n$ has a fixed-dimensional representation with a spike-and-slab prior on the flux.
However, standard MCMCs are not able to explore a posterior that is specified in terms of Dirac deltas.
We can solve this problem with a flux reparameterization.
Let $p(F_i)$ be the slab-and-spike prior of \eqref{eq:binom_factored}:
\begin{equation}
    p(F_i) = \frac{N}{\Nmax} \pi(F_i | \v{\theta}) + \left(1 - \frac{N}{\Nmax}\right) \delta(F_i). \label{eq:ssprior}
\end{equation}
If, instead of sampling in $F_i$ space, we sample in $\sigma_i \in [0, 1]$ where $F_i = f_{N,\v{\theta}}(\sigma_i)$, then we have performed a reparameterization of the fluxes in term of these $\sigma$ variables.
The effect of this reparameterization can be quantified by examining how $p(F_i)$ becomes a pushforward distribution of $p(\sigma_i)$ by $f$:
\begin{equation}
    p(F_i) = \int_0^1 \d \sigma_i p(\sigma_i) \delta(F_i - f_{N,\v{\theta}}(\sigma_i)). \label{eq:pushforward}
\end{equation}
We need to choose $f$ and $p(\sigma_i)$ such that the equality of \eqref{eq:ssprior} and \eqref{eq:pushforward} holds. 

To generate the two terms in \eqref{eq:ssprior}, we may split this integral into two parts by choosing $f$ to be a piecewise function as shown in figure~\ref{fig:sspriorregular} (middle):
\begin{equation}
    f_{N,\v{\theta}}(\sigma_i) = \begin{cases} \CDF^{-1}_{\pi(F_i|\v{\theta})}\left(1 - \frac{\sigma_i}{N/\Nmax}\right) & \sigma_i < \frac{N}{\Nmax} \\ 0 & \text{otherwise} \end{cases},
\end{equation}
where $\CDF^{-1}_{\pi(F_i|\v{\theta})}$ is the inverse cumulative distribution function (CDF) for the prior $\pi(F_i|\v{\theta})$.
The choice of $1 - \frac{\sigma_i}{N/\Nmax}$ as the argument to the inverse CDF ensures that $f$ is continuous at $\sigma_i = \frac{N}{\Nmax}$, since we require that $\CDF^{-1}_{\pi(F_i|\v{\theta})}(0) = 0$.

The second (spike) term in \eqref{eq:ssprior} is generated by the second piece of this function --- we simply need
\begin{equation}
    \left(1 - \frac{N}{\Nmax}\right) \delta(F_i) = \int_{N/\Nmax}^1 \d \sigma_i p(\sigma_i) \delta(F_i - f_{N,\v{\theta}}(\sigma_i)),
\end{equation}
which holds if $p(\sigma_i)$ is a uniform distribution.
The first (slab) term in \eqref{eq:ssprior} is generated by the first piece of $f$.
This piece is bijective, so we can use the usual Jacobian change of variables identity to recover the slab:
\begin{equation}
    \delta(F_i - f_{N,\v{\theta}}(\sigma_i)) = \delta(\sigma_i - f^{-1}_{N,\v{\theta}}(F_i)) \left| \frac{d f^{-1}_{N,\v{\theta}}(F_i)}{d F_i} \right|,
\end{equation}
where $f^{-1}_{N,\v{\theta}}$ is the inverse of the bijective part of $f$:
\begin{equation}
    f^{-1}_{N,\v{\theta}}(F_i) = \frac{N}{\Nmax} \left[1 - \CDF_{\pi(F_i|\v{\theta})}(F_i)\right].
\end{equation}
As the derivative of the CDF is the flux prior $\pi(F_i|\v{\theta})$, the 
equality of \eqref{eq:ssprior} and \eqref{eq:pushforward} holds.
The transformation from $\sigma_i$ through $f$ to generate \eqref{eq:ssprior} is illustrated in figure~\ref{fig:sspriorregular}.



Now that we have shown that reparameterization by $f$ generates the slab-and-spike prior, we can perform the change of variables in \eqref{eq:binom_factored} to
\begin{multline}
    \lh(\v{d}|N,\v{\theta}) = \left(\prod_{i=1}^{\Nmax}  \int \d\v{\phi}_i \pi(\v{\phi}_i|\v{\theta}) \int \d \sigma_i p(\sigma_i) \right) \\ \lh(\v{d}|\Nmax,\{\v{\phi}_i\},\{f_{N,\v{\theta}}(\sigma_i)\}). \label{eq:likelihood}
\end{multline}
Therefore if we run an MCMC in the joint and fixed-dimensional space of $\{\v{\phi}_i\}$, $\{\sigma_i\}$, $N$, and $\v{\theta}$, then we can perform probabilistic cataloging with a binomial prior on $n$. 
The value of $n$ for any sample from the posterior can be calculated by counting the number of non-zero flux coordinates.

\subsection{Sampling}

The state space of the MCMC will typically contain three or more parameters per source.
For example, $\Nmax \sim 10^{4.5}$ requires $\sim 10^5$ parameters.
Microcanonical Langevin Monte Carlo (MCLMC) \cite{10.5555/3648699.3649010,robnik2024fluctuation} is a recent statistical sampler that has been shown to scale to problems with $10^5$ parameters \cite{Bayer:2023rmj}.
For smaller $\Nmax$ it may be possible to employ HMC.
Here MCLMC as implemented in the \texttt{blackjax} library \cite{cabezasBlackJAXComposableBayesian2024} is used.

\begin{figure}
  \centering
  \includegraphics[page=2,width=0.8\columnwidth]{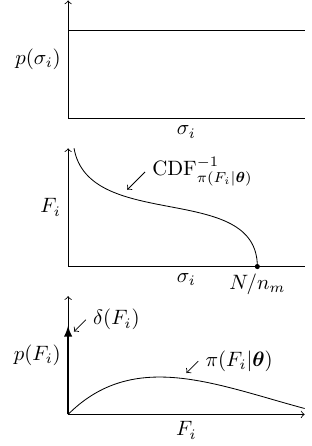}
  \caption{As figure~\ref{fig:sspriorregular} in the specific case that $\pi(F|\v{\theta}) \to 0$ as $F \to 0$ (shown at bottom). Middle: The derivative of $f$ now diverges as $\sigma_i \to N/\Nmax$.} 
  \label{fig:sspriorinf}
\end{figure}

Both MCLMC and HMC are gradient based samplers which are best behaved when both the target distribution and its derivative are continuous.
The parametric cataloging likelihood is piecewise continuous, but its derivative is not. 
As the sampler trajectory crosses the discontinuity it picks up an integration error that leads to a bias in the final sample.
If necessary, this bias can be removed at the expense of additional variance through a Metropolis-Hastings correction step.
However, even with this correction there still exists a pathology if $\pi(F|\v{\theta}) \to 0$ as $F \to 0$.
In this case, the derivative of the CDF also approaches zero at $F \to 0$ and thus the derivative of the inverse CDF diverges to infinity at the corresponding point leading to the derivative of the likelihood diverging as any $\sigma_i \to N/\Nmax$.
This is illustrated in figure~\ref{fig:sspriorinf}.
For large $\Nmax$ it is highly probable that there will be at least one source for which $\sigma_i$ is close enough to $N/\Nmax$ to produce a derivative of the likelihood that is large enough to all but guarantee either a large sampling bias or that the proposed trajectories are rejected by the Metropolis-Hastings correction.

A restriction on $\pi(F|\v{\theta})$ will allow us to control the discontinuity in the derivative of the likelihood.
Suppose $\pi(F=0|\v{\theta}) > 0$, then the derivative of the likelihood will be no worse than a constant in the neighborhood of $\sigma_i = N/\Nmax$.
A common choice for $\pi(F|\v{\theta})$ is the singly-broken power law.
If the final (lowest flux) index, $n_2$, has the typical domain of $n_2 \leq 0$, then for all $n_f < 0$, $\pi(F=0|\v{\theta}) = 0$, exhibiting the pathological behavior described above.
Thus, it is prudent to control this pathology by adopting a doubly-broken power law where the final index, $n_3$, is pinned to zero while $n_2 \leq 0$.
The result is that $\pi(F=0|\v{\theta}) > 0$.

\section{Demonstration}

To validate this framework, a demonstration has been constructed that broadly follows \textcite{Collin:2021ufc}.
This demonstration may be downloaded from \url{https://github.com/ghcollin/paracat}.
In particular, an X-ray telescope similar to NuSTAR \cite{Wik:2014boa,Madsen:2015jea,madsenObservationalArtifactsNuSTAR2017} is simulated with a 12 arcmin field of view, a $64\times64$ pixelated detector, producing four non-overlapping contiguous observations.
A Gaussian Point Spread Function (PSF) with a radius of 0.25 arcmin and a Gaussian effective area with radius of 20 arcmin.
The scale of the effective area, and a uniform background of 4 counts per pixel, were chosen to reflect an observation of 200 ks.
The Gaussian PSF and effective area are not representative of NuSTAR, but are significantly faster to evaluate on a GPU.
The NuSTAR PSF and effective area are specified by the CalDB \cite{NuSTAR:2013yza} as arrays that are used as look up tables.
This significantly increases GPU memory usage and requires random memory access that slows the algorithm by an order of magnitude.
No ghost rays or stray light are simulated.
The expected counts for each source is evaluated in $12\times12$ sub-images which are then composited into the final $64\times64$ observations for evaluation of the Poisson likelihood.
A choice of $\Nmax=40,000$ was selected as it represents close to the maximum possible that fits in the 24GB of VRAM available on the NVIDIA RTX 4090 used for this demonstration.
This limit is caused by the reverse-mode automatic differentiation required for the gradient evaluation in the sampler, and is not fundamental to the Parametric Cataloging framework.

\begin{figure*}
  \centering
  \includegraphics[page=1,width=2\columnwidth]{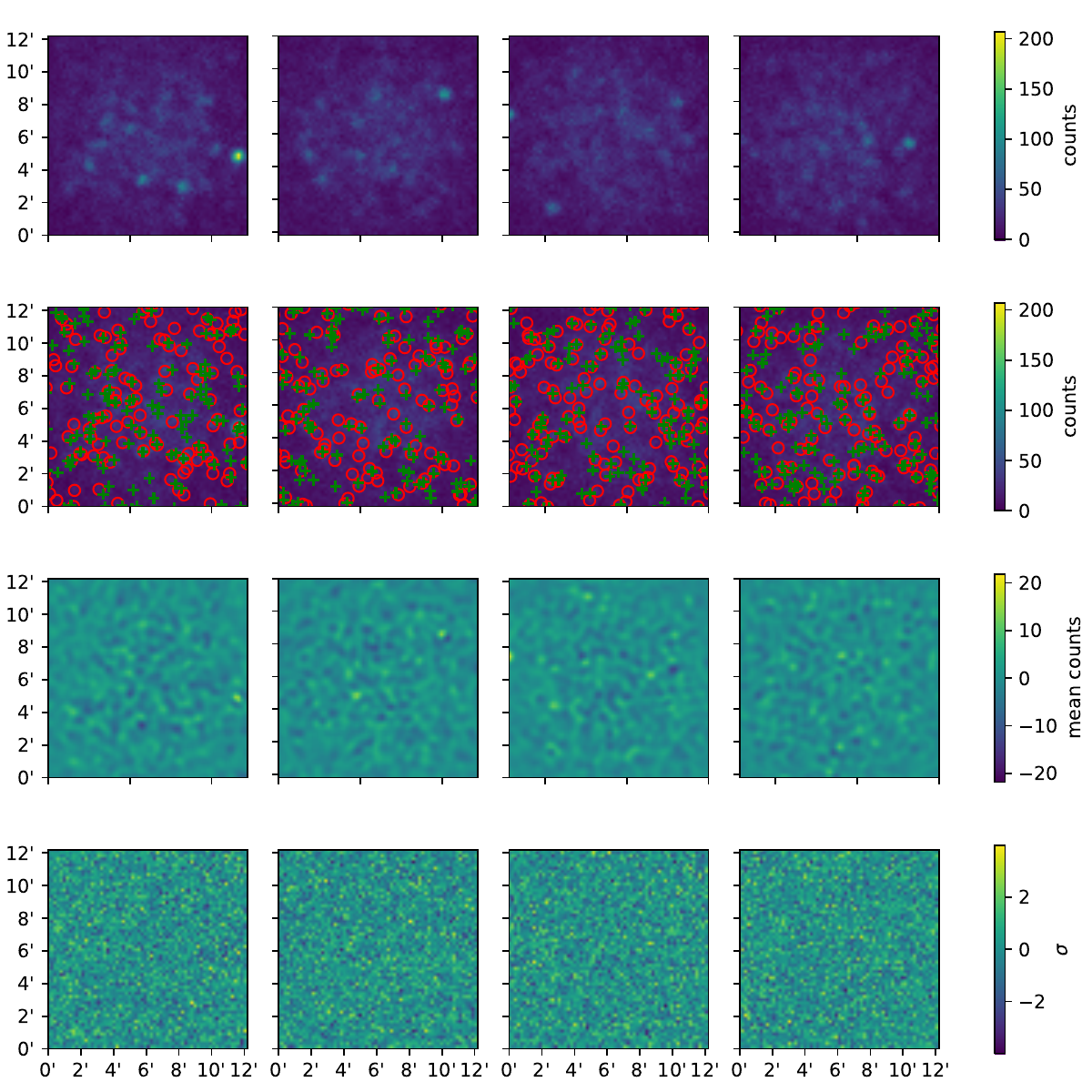}
  \caption{Left to right shows the four observations, see the demonstration section for interpretation. Top: Each of the simulated observations, while some of the brightest sources are resolvable, the vast majority of the $\sim$3,200 sources in each of these images are not. Second: Red circles show the true locations of sources that contribute 100 counts or more, green pluses show locations of putative sources from one sample catalog of the posterior. Third: The expected counts from this sample catalog is subtracted to show residuals. Bottom: The statistical significance of these residuals is shown in $\sigma$-values.}
  \label{fig:example_obs}
\end{figure*}

Sources are injected into the image according to a doubly-broken power law with indices $n_1,n_2,n_3=3,-2,0$, source density of $2 K$ arcmin${}^{-1}$, and flux breaks $F_{b,1} = K^{-1/n_1} 6 \times 10^{-9}$ s${}^{-1}$ cm${}^{-2}$, $F_{b,2}=0.1 F_{b,2}$.
The parameter $K$ is used to increase the number of sources, while ensuring the number of sources at high flux remains constant.
A value of $K=1$ corresponds to the ``realistic scenario'' in \textcite{Collin:2021ufc}, without a Cosmic X-Ray Background as this diffuse flux component tends to complicate round-trip recovery of $N$.
A value of $K=10$ was chosen for this demonstration, which corresponds to approximately 27,000 total sources in the population.
The spatial distribution of the injected sources is uniform and extends outside the imaged area by $25\%$ on each side, thus approximately 3,200 sources are within the field of view of each of the four observations.
The expected number of counts from each source is computed for each pixel using the PSF and effective area, then added to the background of 4 counts forming a mean counts per pixel from which a Poisson distributed value was drawn to create the simulated image.
These four observations are shown on the top row of figure~\ref{fig:example_obs}.

During inference, $\pi(F|\v{\theta})$ is specified as a doubly-broken power law in the natural coordinates of \textcite{Collin:2021ufc}. 
The priors on $n_1$ and $n_2$ are uniform in angle-space as advocated in \textcite{Collin:2021ufc} and span $n_1 \in [2.1, 4]$, $n_2 \in [-4, 0]$ with $n_3 = 0$ fixed.
The prior on the total flux, $F_T$, is uniform in log-space and spans $F_T \in [-5, -3.5]$, while the prior on the flux fraction parameter $\beta$ is uniform in $\beta \in [0.03, 0.3]$.
The prior on the mean number of sources, $N$, is uniform over $N \in [1, \Nmax-1]$.
The prior on the spatial distribution of the sources is uniform and matches the distribution used to inject the sources.
The prior on the background is log-normal with mean 4 and standard deviation 0.2.

At the beginning of inference, the parameters are drawn randomly from their priors.
Recall that the sources will not be parameterized in terms of their fluxes, $F_i$, but by the auxiliary variable $\sigma_i$ instead, and the prior for these is the unit uniform distribution.
The likelihood is expected to be highly multimodal, and the sampler may get trapped at low $N$.
To combat this, $N$ is not initialized to the prior, and is instead initialized at the 99.865\textsuperscript{th} percentile (3 standard deviations).

\begin{figure}
  \centering
  \includegraphics[page=5,width=\columnwidth]{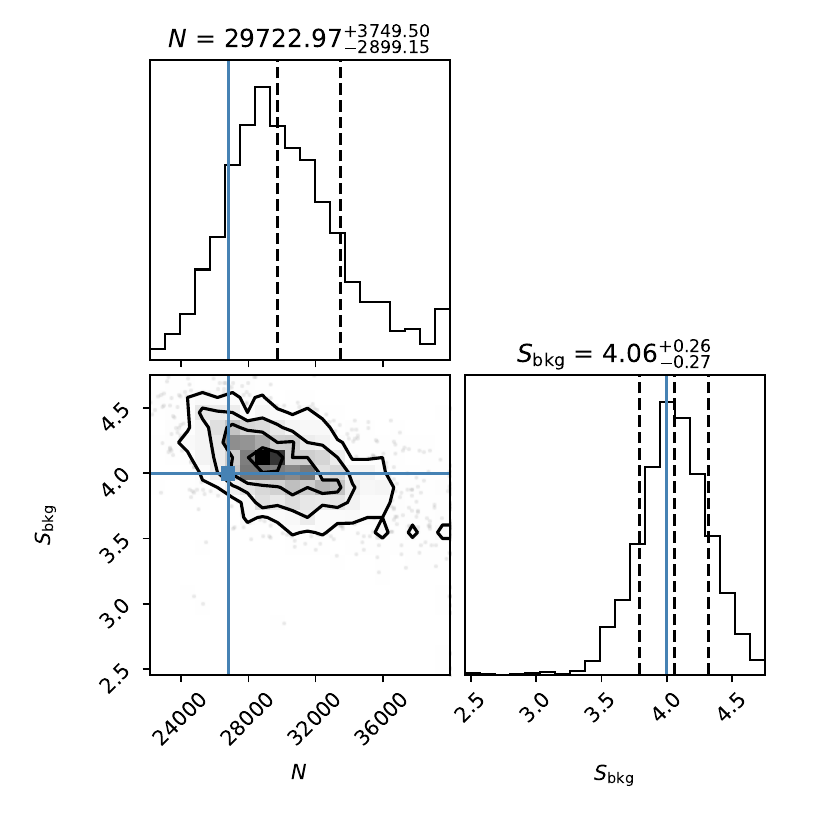}
  \includegraphics[page=7,width=\columnwidth]{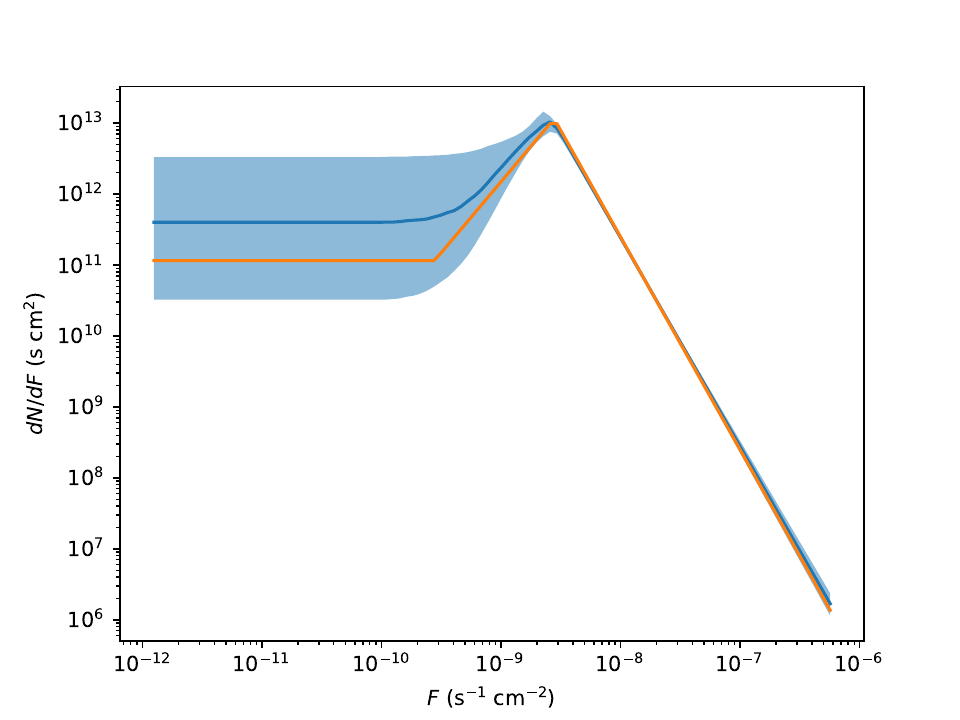}
  \caption{Above: The posterior distribution over the mean sources in the population, $N$, and the parameterization of the background component, $S_\textrm{bkg}$. The 16\%, median, and 84\% quantiles are shown in dashed black lines while the true parameters are shown in blue lines. Below: The injected flux spectrum is shown in orange while the posterior of the flux spectrum is shown in blue. The blue line shows the median while the blue bands show the 16\% and 84\% quantiles.}
  \label{fig:posterior}
\end{figure}

The unadjusted MCLMC algorithm as implemented in \texttt{blackjax} uses the autocorrelation of a warm-up sample to estimate the correct momentum decoherence scale hyperparameter.
The autocorrelation estimator uses a fast Fourier transform that requires high VRAM usage on this exceptionally high-dimensional inference problem.
Therefore, only $0.5\%$ of the warm-up stage is used to estimate the momentum decoherence scale.
The Metropolis-Hastings-correction adjusted version may be used in scenarios where this scale cannot be reliably estimated from a small subset of the warm-up stage.
This version is closer to HMC in operation, and uses an on-line dual averaging scheme to estimate the step size.

Each chain contains $10^6$ samples, thinned by a factor of $10^3$, with a $5\times10^5$ sample warm-up, leading to a run time of 2 GPU-hours.
This corresponds to $\mathcal{O}(10^6)$ likelihood evaluations, which compares well to the $\mathcal{O}(10^6)$ full-likelihood-equivalent evaluations required in \textcite{Daylan:2016tia} for a problem with two orders of magnitude fewer sources.
The posterior distribution for $N$ and the background parameter $S_\textrm{bkg}$ is shown in figure~\ref{fig:posterior} along with the posterior of the flux spectrum.
The posterior recovers the true parameters well, with a slight over-estimation on the number of sources.
This over-estimation is highly prior driven due to the inherent ambiguity caused by the large number of low flux sources.
The posterior on $N$ also shows a slight pile-up at the upper edge of the prior caused by the bias due to integration errors accumulated by the sampler. 
This may be controlled by reducing the step size, or switching to the Metropolis-Hastings-corrected version of the sampler at the expense of significant additional computation time.

One sample is chosen from the posterior and used to generate the three bottom rows of figure~\ref{fig:example_obs}, where each column corresponds to one of the four observations.
Sources that contribute 100 counts or more are shown on the second row, with true source locations in red circles and the locations of putative sources in the posterior same shown in green crosses.
Note that these sources contribute only a small fraction of the total flux in the image, and so many true source locations do not have nearby putative sources as their flux may be provided by dim sources below the plotting threshold.
To gain a better view on the quality of the sample, we can plot the residuals, as shown in the third row.
Residuals show fluctuations on the order of 10 counts uniformly distributed over the view, which are the expected statistical fluctuations.
We can see this by examining the bottom row, which shows the statistical significance of the fluctuations as $\sigma$-values.
The significances are also uniformly distributed over the view, with no hot spots corresponding to poorly fitted regions.

The fixed-dimensional nature of the algorithm permits the standard convergence metrics of MCMC to be used.
The Potential Scale Reduction Factor (PSRF, $\hat{R}$) is calculated for all parameters using two independent chains.
For the model parameters, $N$ and $\v{\theta}$, the corresponding $\hat{R}$ were found to be less than 1.05 with greater than 400 effective samples.
It is also possible to calculate $\hat{R}$ for the flux and location parameters of the sources, but with so many such parameters it is statistically likely to find examples in any otherwise well-converged chain that have large $\hat{R}$.
Instead, we can compute the 99\textsuperscript{th} percentile which were all $<1.05$ for the fluxes and each location coordinate.

\subsection{Effect of binomial prior} \label{sec:binom}

The major methodological caveat of Parametric Cataloging is the need to use a binomial prior on the number of sources ($n$).
In the typical scientific scenario, the maximum number of sources ($\Nmax$) is not known, and so a Poisson distribution is the appropriate choice.
In comparison, a binomial distribution is more informative and may lead to small posterior distributions on the mean sources in the population ($N$).
We can test this effect by considering two scenarios: one with $\Nmax=4,000$ and $K=1$ ($\sim$2,700 injected sources), effectively a smaller scale version of the above demonstration; and one with $\Nmax=40,000$ but also with $K=1$, such that the number of true sources is less than $10\Nmax$ leading to an inferred $N$ such that $N/\Nmax < 0.1$.
In the $N/\Nmax < 0.1$ regime, the binomial distribution will behave approximately as a Poisson distribution; therefore, the two scenarios under consideration are effectively a binomial prior and an approximate Poisson prior.
These tests recover posterior distributions over $N$ with median and 16\%/84\% quantiles of $2779^{+318}_{-279}$ for the binomial scenario and $2771^{+339}_{-259}$ for the approximate Poisson scenario.
Thus, the difference in the widths of the posterior distributions is minimal and the posteriors are dominated by other effects such as the inherent ambiguity in the data and the prior on $N$.

\section{Discussion}

This demonstration shows that Parametric Cataloging is capable of recovering the injected population parameters in a controlled simulation.
In practice, the true number of sources is not known ahead of time, and could be potentially very large.
Probabilistic and Parametric Cataloging cannot describe the $N\rightarrow\infty$ diffuse limit of the population.
Parametric Cataloging has an explicit maximum ($\Nmax$) number of sources, while probabilistic cataloging has a practical upper limit based on available computational resources.
Both frameworks may capture the $N\rightarrow\infty$ diffuse limit to a good approximation by including a diffuse flux template that follows the spatial prior of the sources.
This is permitted by the inherent degeneracy between a diffuse flux and a population of dim and numerous sources \cite{Collin:2021ufc}.
In Parametric Cataloging, a guess of $\Nmax$ must be made ahead of time such that $\Nmax$ is larger than the remaining bright sources.
A guess that is too low will be revealed by a posterior that converges to the edge of the prior; while, as shown in section~\ref{sec:binom}, there is no penalty (other than computational expense) for choosing a $\Nmax$ that is much larger than the true number of sources, and may in fact be beneficial if a Poisson prior on the number of sources is desired.

Astronomical data also often includes energy information of arriving particles or multi-band images.
This information may be incorporated into Parametric Cataloging through an appropriate choice of likelihood.
For example: we may incorporate an energy spectrum whose overall flux is controlled by $F_i$ with the addition of any necessary additional per-source or population-level parameters such as spectral indices to control the shape.
With this, the likelihood may be computed in an un-binned manner or through a Poisson distribution of expected-count images generated for each energy bin or spectral band.

\section{Conclusion}

Parametric Cataloging provides an accessible and efficient framework for performing point-source inference.
As a manifestation of probabilistic cataloging, it provides a full posterior distribution over catalogs. 
The assumption of a binomial prior on $n$ allows probabilistic cataloging to be embedded in a fixed-dimensional framework that permits high-performance gradient-based samplers such as the recent MCLMC sampler.
This permits probabilistic cataloging with more than an order of magnitude more sources compared with previous approaches, while using comparable computational resources.
Round-trip recovery of population-level parameters has been demonstrated in a NuSTAR-esque scenario.
Parametric Cataloging can be implemented by writing a likelihood model for a fixed number of sources, and then extending this to a variable number of sources by introducing the flux reparameterization to $\sigma_i$.
Transitions between numbers of sources, which must be implemented manually in probabilistic cataloging, are then handled automatically by the sampler.

\section{Acknowledgments}

The author would like to thank Gary Hill for discussions on this topic; Richard Feder, Douglas Finkbeiner, and Nick Rodd for constructive comments on a draft of this work; and the Ramsay Fellowship for financial support during the development of this method.

\bibliography{bib.bib}

\begin{thebibliography}{47}%
\makeatletter
\providecommand \@ifxundefined [1]{%
 \@ifx{#1\undefined}
}%
\providecommand \@ifnum [1]{%
 \ifnum #1\expandafter \@firstoftwo
 \else \expandafter \@secondoftwo
 \fi
}%
\providecommand \@ifx [1]{%
 \ifx #1\expandafter \@firstoftwo
 \else \expandafter \@secondoftwo
 \fi
}%
\providecommand \natexlab [1]{#1}%
\providecommand \enquote  [1]{``#1''}%
\providecommand \bibnamefont  [1]{#1}%
\providecommand \bibfnamefont [1]{#1}%
\providecommand \citenamefont [1]{#1}%
\providecommand \href@noop [0]{\@secondoftwo}%
\providecommand \href [0]{\begingroup \@sanitize@url \@href}%
\providecommand \@href[1]{\@@startlink{#1}\@@href}%
\providecommand \@@href[1]{\endgroup#1\@@endlink}%
\providecommand \@sanitize@url [0]{\catcode `\\12\catcode `\$12\catcode `\&12\catcode `\#12\catcode `\^12\catcode `\_12\catcode `\%12\relax}%
\providecommand \@@startlink[1]{}%
\providecommand \@@endlink[0]{}%
\providecommand \url  [0]{\begingroup\@sanitize@url \@url }%
\providecommand \@url [1]{\endgroup\@href {#1}{\urlprefix }}%
\providecommand \urlprefix  [0]{URL }%
\providecommand \Eprint [0]{\href }%
\providecommand \doibase [0]{https://doi.org/}%
\providecommand \selectlanguage [0]{\@gobble}%
\providecommand \bibinfo  [0]{\@secondoftwo}%
\providecommand \bibfield  [0]{\@secondoftwo}%
\providecommand \translation [1]{[#1]}%
\providecommand \BibitemOpen [0]{}%
\providecommand \bibitemStop [0]{}%
\providecommand \bibitemNoStop [0]{.\EOS\space}%
\providecommand \EOS [0]{\spacefactor3000\relax}%
\providecommand \BibitemShut  [1]{\csname bibitem#1\endcsname}%
\let\auto@bib@innerbib\@empty
\bibitem [{\citenamefont {Hogg}\ and\ \citenamefont {Lang}(2010)}]{Hogg:2010ip}%
  \BibitemOpen
  \bibfield  {author} {\bibinfo {author} {\bibfnamefont {D.}~\bibnamefont {Hogg}}\ and\ \bibinfo {author} {\bibfnamefont {D.}~\bibnamefont {Lang}},\ }\bibfield  {title} {\bibinfo {title} {Telescopes don't make catalogues!},\ }\href {https://doi.org/10.1051/eas/1045059} {\bibfield  {journal} {\bibinfo  {journal} {EAS Publications Series}\ }\textbf {\bibinfo {volume} {45}},\ \bibinfo {pages} {351} (\bibinfo {year} {2010})}\BibitemShut {NoStop}%
\bibitem [{\citenamefont {Brewer}\ \emph {et~al.}(2013)\citenamefont {Brewer}, \citenamefont {{Foreman-Mackey}},\ and\ \citenamefont {Hogg}}]{Brewer:2012gt}%
  \BibitemOpen
  \bibfield  {author} {\bibinfo {author} {\bibfnamefont {B.~J.}\ \bibnamefont {Brewer}}, \bibinfo {author} {\bibfnamefont {D.}~\bibnamefont {{Foreman-Mackey}}},\ and\ \bibinfo {author} {\bibfnamefont {D.~W.}\ \bibnamefont {Hogg}},\ }\bibfield  {title} {\bibinfo {title} {{{PROBABILISTIC CATALOGS FOR CROWDED STELLAR FIELDS}}},\ }\href {https://doi.org/10.1088/0004-6256/146/1/7} {\bibfield  {journal} {\bibinfo  {journal} {The Astronomical Journal}\ }\textbf {\bibinfo {volume} {146}},\ \bibinfo {pages} {7} (\bibinfo {year} {2013})}\BibitemShut {NoStop}%
\bibitem [{\citenamefont {Daylan}\ \emph {et~al.}(2017)\citenamefont {Daylan}, \citenamefont {Portillo},\ and\ \citenamefont {Finkbeiner}}]{Daylan:2016tia}%
  \BibitemOpen
  \bibfield  {author} {\bibinfo {author} {\bibfnamefont {T.}~\bibnamefont {Daylan}}, \bibinfo {author} {\bibfnamefont {S.~K.~N.}\ \bibnamefont {Portillo}},\ and\ \bibinfo {author} {\bibfnamefont {D.~P.}\ \bibnamefont {Finkbeiner}},\ }\bibfield  {title} {\bibinfo {title} {Inference of {{Unresolved Point Sources}} at {{High Galactic Latitudes Using Probabilistic Catalogs}}},\ }\href {https://doi.org/10.3847/1538-4357/aa679e} {\bibfield  {journal} {\bibinfo  {journal} {The Astrophysical Journal}\ }\textbf {\bibinfo {volume} {839}},\ \bibinfo {pages} {4} (\bibinfo {year} {2017})}\BibitemShut {NoStop}%
\bibitem [{\citenamefont {Green}(1995)}]{Green:1995mxx}%
  \BibitemOpen
  \bibfield  {author} {\bibinfo {author} {\bibfnamefont {P.~J.}\ \bibnamefont {Green}},\ }\bibfield  {title} {\bibinfo {title} {Reversible jump {{Markov}} chain {{Monte Carlo}} computation and {{Bayesian}} model determination},\ }\href {https://doi.org/10.1093/biomet/82.4.711} {\bibfield  {journal} {\bibinfo  {journal} {Biometrika}\ }\textbf {\bibinfo {volume} {82}},\ \bibinfo {pages} {711} (\bibinfo {year} {1995})}\BibitemShut {NoStop}%
\bibitem [{\citenamefont {Richardson}\ and\ \citenamefont {Green}(1997)}]{richardsonBayesianAnalysisMixtures1997a}%
  \BibitemOpen
  \bibfield  {author} {\bibinfo {author} {\bibfnamefont {{\relax Sylvia}.}~\bibnamefont {Richardson}}\ and\ \bibinfo {author} {\bibfnamefont {P.~J.}\ \bibnamefont {Green}},\ }\bibfield  {title} {\bibinfo {title} {On {{Bayesian Analysis}} of {{Mixtures}} with an {{Unknown Number}} of {{Components}} (with discussion)},\ }\href {https://doi.org/10.1111/1467-9868.00095} {\bibfield  {journal} {\bibinfo  {journal} {Journal of the Royal Statistical Society Series B: Statistical Methodology}\ }\textbf {\bibinfo {volume} {59}},\ \bibinfo {pages} {731} (\bibinfo {year} {1997})}\BibitemShut {NoStop}%
\bibitem [{\citenamefont {Brewer}\ \emph {et~al.}(2016)\citenamefont {Brewer}, \citenamefont {Huijser},\ and\ \citenamefont {Lewis}}]{Brewer:2015yya}%
  \BibitemOpen
  \bibfield  {author} {\bibinfo {author} {\bibfnamefont {B.~J.}\ \bibnamefont {Brewer}}, \bibinfo {author} {\bibfnamefont {D.}~\bibnamefont {Huijser}},\ and\ \bibinfo {author} {\bibfnamefont {G.~F.}\ \bibnamefont {Lewis}},\ }\bibfield  {title} {\bibinfo {title} {Trans-dimensional {{Bayesian}} inference for gravitational lens substructures},\ }\href {https://doi.org/10.1093/mnras/stv2370} {\bibfield  {journal} {\bibinfo  {journal} {Monthly Notices of the Royal Astronomical Society}\ }\textbf {\bibinfo {volume} {455}},\ \bibinfo {pages} {1819} (\bibinfo {year} {2016})}\BibitemShut {NoStop}%
\bibitem [{\citenamefont {Jones}\ \emph {et~al.}(2015)\citenamefont {Jones}, \citenamefont {Kashyap},\ and\ \citenamefont {Van~Dyk}}]{jonesDISENTANGLINGOVERLAPPINGASTRONOMICAL2015}%
  \BibitemOpen
  \bibfield  {author} {\bibinfo {author} {\bibfnamefont {D.~E.}\ \bibnamefont {Jones}}, \bibinfo {author} {\bibfnamefont {V.~L.}\ \bibnamefont {Kashyap}},\ and\ \bibinfo {author} {\bibfnamefont {D.~A.}\ \bibnamefont {Van~Dyk}},\ }\bibfield  {title} {\bibinfo {title} {{{DISENTANGLING OVERLAPPING ASTRONOMICAL SOURCES USING SPATIAL AND SPECTRAL INFORMATION}}},\ }\href {https://doi.org/10.1088/0004-637X/808/2/137} {\bibfield  {journal} {\bibinfo  {journal} {The Astrophysical Journal}\ }\textbf {\bibinfo {volume} {808}},\ \bibinfo {pages} {137} (\bibinfo {year} {2015})}\BibitemShut {NoStop}%
\bibitem [{\citenamefont {Brewer}\ and\ \citenamefont {Donovan}(2015)}]{brewerFastBayesianInference2015}%
  \BibitemOpen
  \bibfield  {author} {\bibinfo {author} {\bibfnamefont {B.~J.}\ \bibnamefont {Brewer}}\ and\ \bibinfo {author} {\bibfnamefont {C.~P.}\ \bibnamefont {Donovan}},\ }\bibfield  {title} {\bibinfo {title} {Fast {{Bayesian}} inference for exoplanet discovery in radial velocity data},\ }\href {https://doi.org/10.1093/mnras/stv199} {\bibfield  {journal} {\bibinfo  {journal} {Monthly Notices of the Royal Astronomical Society}\ }\textbf {\bibinfo {volume} {448}},\ \bibinfo {pages} {3206} (\bibinfo {year} {2015})}\BibitemShut {NoStop}%
\bibitem [{\citenamefont {Portillo}\ \emph {et~al.}(2017)\citenamefont {Portillo}, \citenamefont {Lee}, \citenamefont {Daylan},\ and\ \citenamefont {Finkbeiner}}]{portilloImprovedPointsourceDetection2017}%
  \BibitemOpen
  \bibfield  {author} {\bibinfo {author} {\bibfnamefont {S.~K.~N.}\ \bibnamefont {Portillo}}, \bibinfo {author} {\bibfnamefont {B.~C.~G.}\ \bibnamefont {Lee}}, \bibinfo {author} {\bibfnamefont {T.}~\bibnamefont {Daylan}},\ and\ \bibinfo {author} {\bibfnamefont {D.~P.}\ \bibnamefont {Finkbeiner}},\ }\bibfield  {title} {\bibinfo {title} {Improved {{Point-source Detection}} in {{Crowded Fields Using Probabilistic Cataloging}}},\ }\href {https://doi.org/10.3847/1538-3881/aa8565} {\bibfield  {journal} {\bibinfo  {journal} {The Astronomical Journal}\ }\textbf {\bibinfo {volume} {154}},\ \bibinfo {pages} {132} (\bibinfo {year} {2017})}\BibitemShut {NoStop}%
\bibitem [{\citenamefont {Daylan}\ \emph {et~al.}(2018)\citenamefont {Daylan}, \citenamefont {{Cyr-Racine}}, \citenamefont {Rivero}, \citenamefont {Dvorkin},\ and\ \citenamefont {Finkbeiner}}]{Daylan:2017kfh}%
  \BibitemOpen
  \bibfield  {author} {\bibinfo {author} {\bibfnamefont {T.}~\bibnamefont {Daylan}}, \bibinfo {author} {\bibfnamefont {F.-Y.}\ \bibnamefont {{Cyr-Racine}}}, \bibinfo {author} {\bibfnamefont {A.~D.}\ \bibnamefont {Rivero}}, \bibinfo {author} {\bibfnamefont {C.}~\bibnamefont {Dvorkin}},\ and\ \bibinfo {author} {\bibfnamefont {D.~P.}\ \bibnamefont {Finkbeiner}},\ }\bibfield  {title} {\bibinfo {title} {Probing the {{Small-scale Structure}} in {{Strongly Lensed Systems}} via {{Transdimensional Inference}}},\ }\href {https://doi.org/10.3847/1538-4357/aaaa1e} {\bibfield  {journal} {\bibinfo  {journal} {The Astrophysical Journal}\ }\textbf {\bibinfo {volume} {854}},\ \bibinfo {pages} {141} (\bibinfo {year} {2018})}\BibitemShut {NoStop}%
\bibitem [{\citenamefont {Sottosanti}\ \emph {et~al.}(2019)\citenamefont {Sottosanti}, \citenamefont {Costantin}, \citenamefont {Bastieri},\ and\ \citenamefont {Brazzale}}]{sottosantiDiscoveringLocatingHighEnergy2019}%
  \BibitemOpen
  \bibfield  {author} {\bibinfo {author} {\bibfnamefont {A.}~\bibnamefont {Sottosanti}}, \bibinfo {author} {\bibfnamefont {D.}~\bibnamefont {Costantin}}, \bibinfo {author} {\bibfnamefont {D.}~\bibnamefont {Bastieri}},\ and\ \bibinfo {author} {\bibfnamefont {A.~R.}\ \bibnamefont {Brazzale}},\ }\bibfield  {title} {\bibinfo {title} {Discovering and {{Locating High-Energy Extra-galactic Sources}} by {{Bayesian Mixture Modelling}}},\ }in\ \href {https://doi.org/10.1007/978-3-030-21158-5_11} {\emph {\bibinfo {booktitle} {New {{Statistical Developments}} in {{Data Science}}}}},\ Vol.\ \bibinfo {volume} {288},\ \bibinfo {editor} {edited by\ \bibinfo {editor} {\bibfnamefont {A.}~\bibnamefont {Petrucci}}, \bibinfo {editor} {\bibfnamefont {F.}~\bibnamefont {Racioppi}},\ and\ \bibinfo {editor} {\bibfnamefont {R.}~\bibnamefont {Verde}}}\ (\bibinfo  {publisher} {Springer International Publishing},\ \bibinfo {address} {Cham},\ \bibinfo {year} {2019})\ pp.\ \bibinfo {pages} {135--148}\BibitemShut {NoStop}%
\bibitem [{\citenamefont {Feder}\ \emph {et~al.}(2020)\citenamefont {Feder}, \citenamefont {Portillo}, \citenamefont {Daylan},\ and\ \citenamefont {Finkbeiner}}]{federMultibandProbabilisticCataloging2020}%
  \BibitemOpen
  \bibfield  {author} {\bibinfo {author} {\bibfnamefont {R.~M.}\ \bibnamefont {Feder}}, \bibinfo {author} {\bibfnamefont {S.~K.~N.}\ \bibnamefont {Portillo}}, \bibinfo {author} {\bibfnamefont {T.}~\bibnamefont {Daylan}},\ and\ \bibinfo {author} {\bibfnamefont {D.}~\bibnamefont {Finkbeiner}},\ }\bibfield  {title} {\bibinfo {title} {Multiband {{Probabilistic Cataloging}}: {{A Joint Fitting Approach}} to {{Point-source Detection}} and {{Deblending}}},\ }\href {https://doi.org/10.3847/1538-3881/ab74cf} {\bibfield  {journal} {\bibinfo  {journal} {The Astronomical Journal}\ }\textbf {\bibinfo {volume} {159}},\ \bibinfo {pages} {163} (\bibinfo {year} {2020})}\BibitemShut {NoStop}%
\bibitem [{\citenamefont {Butler}\ \emph {et~al.}(2022)\citenamefont {Butler}, \citenamefont {Feder}, \citenamefont {Daylan}, \citenamefont {Mantz}, \citenamefont {Mercado}, \citenamefont {Monta{\~n}a}, \citenamefont {Portillo}, \citenamefont {Sayers}, \citenamefont {Vaughan}, \citenamefont {Zemcov},\ and\ \citenamefont {Zitrin}}]{Butler:2021val}%
  \BibitemOpen
  \bibfield  {author} {\bibinfo {author} {\bibfnamefont {V.~L.}\ \bibnamefont {Butler}}, \bibinfo {author} {\bibfnamefont {R.~M.}\ \bibnamefont {Feder}}, \bibinfo {author} {\bibfnamefont {T.}~\bibnamefont {Daylan}}, \bibinfo {author} {\bibfnamefont {A.~B.}\ \bibnamefont {Mantz}}, \bibinfo {author} {\bibfnamefont {D.}~\bibnamefont {Mercado}}, \bibinfo {author} {\bibfnamefont {A.}~\bibnamefont {Monta{\~n}a}}, \bibinfo {author} {\bibfnamefont {S.~K.~N.}\ \bibnamefont {Portillo}}, \bibinfo {author} {\bibfnamefont {J.}~\bibnamefont {Sayers}}, \bibinfo {author} {\bibfnamefont {B.~J.}\ \bibnamefont {Vaughan}}, \bibinfo {author} {\bibfnamefont {M.}~\bibnamefont {Zemcov}},\ and\ \bibinfo {author} {\bibfnamefont {A.}~\bibnamefont {Zitrin}},\ }\bibfield  {title} {\bibinfo {title} {Measurement of the {{Relativistic Sunyaev-Zeldovich Correction}} in {{RX J1347}}.5-1145},\ }\href {https://doi.org/10.3847/1538-4357/ac6c04} {\bibfield  {journal} {\bibinfo  {journal} {The Astrophysical Journal}\ }\textbf {\bibinfo {volume}
  {932}},\ \bibinfo {pages} {55} (\bibinfo {year} {2022})}\BibitemShut {NoStop}%
\bibitem [{\citenamefont {Costantin}\ \emph {et~al.}(2022)\citenamefont {Costantin}, \citenamefont {Sottosanti}, \citenamefont {Brazzale}, \citenamefont {Bastieri},\ and\ \citenamefont {Fan}}]{costantinBayesianMixtureModelling2022}%
  \BibitemOpen
  \bibfield  {author} {\bibinfo {author} {\bibfnamefont {D.}~\bibnamefont {Costantin}}, \bibinfo {author} {\bibfnamefont {A.}~\bibnamefont {Sottosanti}}, \bibinfo {author} {\bibfnamefont {A.~R.}\ \bibnamefont {Brazzale}}, \bibinfo {author} {\bibfnamefont {D.}~\bibnamefont {Bastieri}},\ and\ \bibinfo {author} {\bibfnamefont {J.}~\bibnamefont {Fan}},\ }\bibfield  {title} {\bibinfo {title} {Bayesian mixture modelling of the high-energy photon counts collected by the {{Fermi Large Area Telescope}}},\ }\href {https://doi.org/10.1177/1471082X20947222} {\bibfield  {journal} {\bibinfo  {journal} {Statistical Modelling}\ }\textbf {\bibinfo {volume} {22}},\ \bibinfo {pages} {175} (\bibinfo {year} {2022})}\BibitemShut {NoStop}%
\bibitem [{\citenamefont {Feder}\ \emph {et~al.}(2023)\citenamefont {Feder}, \citenamefont {Butler}, \citenamefont {Daylan}, \citenamefont {Portillo}, \citenamefont {Sayers}, \citenamefont {Vaughan}, \citenamefont {Zamora},\ and\ \citenamefont {Zemcov}}]{federPCATDEReconstructingPointlike2023}%
  \BibitemOpen
  \bibfield  {author} {\bibinfo {author} {\bibfnamefont {R.~M.}\ \bibnamefont {Feder}}, \bibinfo {author} {\bibfnamefont {V.}~\bibnamefont {Butler}}, \bibinfo {author} {\bibfnamefont {T.}~\bibnamefont {Daylan}}, \bibinfo {author} {\bibfnamefont {S.~K.~N.}\ \bibnamefont {Portillo}}, \bibinfo {author} {\bibfnamefont {J.}~\bibnamefont {Sayers}}, \bibinfo {author} {\bibfnamefont {B.~J.}\ \bibnamefont {Vaughan}}, \bibinfo {author} {\bibfnamefont {C.~V.}\ \bibnamefont {Zamora}},\ and\ \bibinfo {author} {\bibfnamefont {M.}~\bibnamefont {Zemcov}},\ }\bibfield  {title} {\bibinfo {title} {{{PCAT-DE}}: {{Reconstructing Pointlike}} and {{Diffuse Signals}} in {{Astronomical Images Using Spatial}} and {{Spectral Information}}},\ }\href {https://doi.org/10.3847/1538-3881/ace69b} {\bibfield  {journal} {\bibinfo  {journal} {The Astronomical Journal}\ }\textbf {\bibinfo {volume} {166}},\ \bibinfo {pages} {98} (\bibinfo {year} {2023})}\BibitemShut {NoStop}%
\bibitem [{\citenamefont {Duane}\ \emph {et~al.}(1987)\citenamefont {Duane}, \citenamefont {Kennedy}, \citenamefont {Pendleton},\ and\ \citenamefont {Roweth}}]{Duane:1987de}%
  \BibitemOpen
  \bibfield  {author} {\bibinfo {author} {\bibfnamefont {S.}~\bibnamefont {Duane}}, \bibinfo {author} {\bibfnamefont {A.}~\bibnamefont {Kennedy}}, \bibinfo {author} {\bibfnamefont {B.~J.}\ \bibnamefont {Pendleton}},\ and\ \bibinfo {author} {\bibfnamefont {D.}~\bibnamefont {Roweth}},\ }\bibfield  {title} {\bibinfo {title} {Hybrid {{Monte Carlo}}},\ }\href {https://doi.org/10.1016/0370-2693(87)91197-X} {\bibfield  {journal} {\bibinfo  {journal} {Physics Letters B}\ }\textbf {\bibinfo {volume} {195}},\ \bibinfo {pages} {216} (\bibinfo {year} {1987})}\BibitemShut {NoStop}%
\bibitem [{\citenamefont {Neal}(2011)}]{Neal:2011mrf}%
  \BibitemOpen
  \bibfield  {author} {\bibinfo {author} {\bibfnamefont {R.}~\bibnamefont {Neal}},\ }\href {https://doi.org/10.1201/b10905} {\emph {\bibinfo {title} {{{MCMC Using Hamiltonian Dynamics}}}}}\ (\bibinfo {year} {2011})\ pp.\ \bibinfo {pages} {113--162}\BibitemShut {NoStop}%
\bibitem [{\citenamefont {Betancourt}(2017)}]{Betancourt:2017ebh}%
  \BibitemOpen
  \bibfield  {author} {\bibinfo {author} {\bibfnamefont {M.}~\bibnamefont {Betancourt}},\ }\href {https://doi.org/10.48550/arXiv.1701.02434} {\bibinfo {title} {A {{Conceptual Introduction}} to {{Hamiltonian Monte Carlo}}}} (\bibinfo {year} {2017})\BibitemShut {NoStop}%
\bibitem [{\citenamefont {Sottosanti}\ \emph {et~al.}(2021)\citenamefont {Sottosanti}, \citenamefont {Bernardi}, \citenamefont {Brazzale}, \citenamefont {{Geringer-Sameth}}, \citenamefont {Stenning}, \citenamefont {Trotta},\ and\ \citenamefont {{van Dyk}}}]{Sottosanti:2021eid}%
  \BibitemOpen
  \bibfield  {author} {\bibinfo {author} {\bibfnamefont {A.}~\bibnamefont {Sottosanti}}, \bibinfo {author} {\bibfnamefont {M.}~\bibnamefont {Bernardi}}, \bibinfo {author} {\bibfnamefont {A.~R.}\ \bibnamefont {Brazzale}}, \bibinfo {author} {\bibfnamefont {A.}~\bibnamefont {{Geringer-Sameth}}}, \bibinfo {author} {\bibfnamefont {D.~C.}\ \bibnamefont {Stenning}}, \bibinfo {author} {\bibfnamefont {R.}~\bibnamefont {Trotta}},\ and\ \bibinfo {author} {\bibfnamefont {D.~A.}\ \bibnamefont {{van Dyk}}},\ }\href {https://doi.org/10.48550/ARXIV.2104.11492} {\bibinfo {title} {Identification of high-energy astrophysical point sources via hierarchical {{Bayesian}} nonparametric clustering}} (\bibinfo {year} {2021})\BibitemShut {NoStop}%
\bibitem [{\citenamefont {Collin}\ \emph {et~al.}(2022)\citenamefont {Collin}, \citenamefont {Rodd}, \citenamefont {Erjavec},\ and\ \citenamefont {Perez}}]{Collin:2021ufc}%
  \BibitemOpen
  \bibfield  {author} {\bibinfo {author} {\bibfnamefont {G.~H.}\ \bibnamefont {Collin}}, \bibinfo {author} {\bibfnamefont {N.~L.}\ \bibnamefont {Rodd}}, \bibinfo {author} {\bibfnamefont {T.}~\bibnamefont {Erjavec}},\ and\ \bibinfo {author} {\bibfnamefont {K.}~\bibnamefont {Perez}},\ }\bibfield  {title} {\bibinfo {title} {A {{Compound Poisson Generator Approach}} to {{Point-source Inference}} in {{Astrophysics}}},\ }\href {https://doi.org/10.3847/1538-4365/ac5cb7} {\bibfield  {journal} {\bibinfo  {journal} {The Astrophysical Journal Supplement Series}\ }\textbf {\bibinfo {volume} {260}},\ \bibinfo {pages} {29} (\bibinfo {year} {2022})}\BibitemShut {NoStop}%
\bibitem [{\citenamefont {Lee}\ \emph {et~al.}(2016)\citenamefont {Lee}, \citenamefont {Lisanti}, \citenamefont {Safdi}, \citenamefont {Slatyer},\ and\ \citenamefont {Xue}}]{Lee:2015fea}%
  \BibitemOpen
  \bibfield  {author} {\bibinfo {author} {\bibfnamefont {S.~K.}\ \bibnamefont {Lee}}, \bibinfo {author} {\bibfnamefont {M.}~\bibnamefont {Lisanti}}, \bibinfo {author} {\bibfnamefont {B.~R.}\ \bibnamefont {Safdi}}, \bibinfo {author} {\bibfnamefont {T.~R.}\ \bibnamefont {Slatyer}},\ and\ \bibinfo {author} {\bibfnamefont {W.}~\bibnamefont {Xue}},\ }\bibfield  {title} {\bibinfo {title} {Evidence for {{Unresolved}} {$\gamma$} -{{Ray Point Sources}} in the {{Inner Galaxy}}},\ }\href {https://doi.org/10.1103/PhysRevLett.116.051103} {\bibfield  {journal} {\bibinfo  {journal} {Physical Review Letters}\ }\textbf {\bibinfo {volume} {116}},\ \bibinfo {pages} {051103} (\bibinfo {year} {2016})}\BibitemShut {NoStop}%
\bibitem [{\citenamefont {Malyshev}\ and\ \citenamefont {Hogg}(2011)}]{Malyshev:2011zi}%
  \BibitemOpen
  \bibfield  {author} {\bibinfo {author} {\bibfnamefont {D.}~\bibnamefont {Malyshev}}\ and\ \bibinfo {author} {\bibfnamefont {D.~W.}\ \bibnamefont {Hogg}},\ }\bibfield  {title} {\bibinfo {title} {Statistics of {{Gamma-Ray Point Sources}} below the {{Fermi Detection Limit}}},\ }\href {https://doi.org/10.1088/0004-637X/738/2/181} {\bibfield  {journal} {\bibinfo  {journal} {The Astrophysical Journal}\ }\textbf {\bibinfo {volume} {738}},\ \bibinfo {pages} {181} (\bibinfo {year} {2011})}\BibitemShut {NoStop}%
\bibitem [{\citenamefont {Lee}\ \emph {et~al.}(2009)\citenamefont {Lee}, \citenamefont {Ando},\ and\ \citenamefont {Kamionkowski}}]{Lee:2008fm}%
  \BibitemOpen
  \bibfield  {author} {\bibinfo {author} {\bibfnamefont {S.~K.}\ \bibnamefont {Lee}}, \bibinfo {author} {\bibfnamefont {S.}~\bibnamefont {Ando}},\ and\ \bibinfo {author} {\bibfnamefont {M.}~\bibnamefont {Kamionkowski}},\ }\bibfield  {title} {\bibinfo {title} {The gamma-ray-flux {{PDF}} from galactic halo substructure},\ }\href {https://doi.org/10.1088/1475-7516/2009/07/007} {\bibfield  {journal} {\bibinfo  {journal} {Journal of Cosmology and Astroparticle Physics}\ }\textbf {\bibinfo {volume} {2009}},\ \bibinfo {pages} {007}}\BibitemShut {NoStop}%
\bibitem [{\citenamefont {Caron}\ \emph {et~al.}(2018)\citenamefont {Caron}, \citenamefont {{G{\'o}mez-Vargas}}, \citenamefont {Hendriks},\ and\ \citenamefont {De~Austri}}]{Caron:2017udl}%
  \BibitemOpen
  \bibfield  {author} {\bibinfo {author} {\bibfnamefont {S.}~\bibnamefont {Caron}}, \bibinfo {author} {\bibfnamefont {G.~A.}\ \bibnamefont {{G{\'o}mez-Vargas}}}, \bibinfo {author} {\bibfnamefont {L.}~\bibnamefont {Hendriks}},\ and\ \bibinfo {author} {\bibfnamefont {R.~R.}\ \bibnamefont {De~Austri}},\ }\bibfield  {title} {\bibinfo {title} {Analyzing {$\gamma$} rays of the {{Galactic Center}} with deep learning},\ }\href {https://doi.org/10.1088/1475-7516/2018/05/058} {\bibfield  {journal} {\bibinfo  {journal} {Journal of Cosmology and Astroparticle Physics}\ }\textbf {\bibinfo {volume} {2018}}\bibinfo  {number} { (05)},\ \bibinfo {pages} {058}}\BibitemShut {NoStop}%
\bibitem [{\citenamefont {List}\ \emph {et~al.}(2020)\citenamefont {List}, \citenamefont {Rodd}, \citenamefont {Lewis},\ and\ \citenamefont {Bhat}}]{List:2020mzd}%
  \BibitemOpen
\bibfield  {number} {  }\bibfield  {author} {\bibinfo {author} {\bibfnamefont {F.}~\bibnamefont {List}}, \bibinfo {author} {\bibfnamefont {N.~L.}\ \bibnamefont {Rodd}}, \bibinfo {author} {\bibfnamefont {G.~F.}\ \bibnamefont {Lewis}},\ and\ \bibinfo {author} {\bibfnamefont {I.}~\bibnamefont {Bhat}},\ }\bibfield  {title} {\bibinfo {title} {Galactic {{Center Excess}} in a {{New Light}}: {{Disentangling}} the {$\gamma$} -{{Ray Sky}} with {{Bayesian Graph Convolutional Neural Networks}}},\ }\href {https://doi.org/10.1103/PhysRevLett.125.241102} {\bibfield  {journal} {\bibinfo  {journal} {Physical Review Letters}\ }\textbf {\bibinfo {volume} {125}},\ \bibinfo {pages} {241102} (\bibinfo {year} {2020})}\BibitemShut {NoStop}%
\bibitem [{\citenamefont {List}\ \emph {et~al.}(2021)\citenamefont {List}, \citenamefont {Rodd},\ and\ \citenamefont {Lewis}}]{List:2021aer}%
  \BibitemOpen
  \bibfield  {author} {\bibinfo {author} {\bibfnamefont {F.}~\bibnamefont {List}}, \bibinfo {author} {\bibfnamefont {N.~L.}\ \bibnamefont {Rodd}},\ and\ \bibinfo {author} {\bibfnamefont {G.~F.}\ \bibnamefont {Lewis}},\ }\bibfield  {title} {\bibinfo {title} {Extracting the {{Galactic Center}} excess' source-count distribution with neural nets},\ }\href {https://doi.org/10.1103/PhysRevD.104.123022} {\bibfield  {journal} {\bibinfo  {journal} {Physical Review D}\ }\textbf {\bibinfo {volume} {104}},\ \bibinfo {pages} {123022} (\bibinfo {year} {2021})}\BibitemShut {NoStop}%
\bibitem [{\citenamefont {{Mishra-Sharma}}\ and\ \citenamefont {Cranmer}(2022)}]{Mishra-Sharma:2021oxe}%
  \BibitemOpen
  \bibfield  {author} {\bibinfo {author} {\bibfnamefont {S.}~\bibnamefont {{Mishra-Sharma}}}\ and\ \bibinfo {author} {\bibfnamefont {K.}~\bibnamefont {Cranmer}},\ }\bibfield  {title} {\bibinfo {title} {Neural simulation-based inference approach for characterizing the {{Galactic Center}} {$\gamma$} -ray excess},\ }\href {https://doi.org/10.1103/PhysRevD.105.063017} {\bibfield  {journal} {\bibinfo  {journal} {Physical Review D}\ }\textbf {\bibinfo {volume} {105}},\ \bibinfo {pages} {063017} (\bibinfo {year} {2022})}\BibitemShut {NoStop}%
\bibitem [{\citenamefont {Caron}\ \emph {et~al.}(2023)\citenamefont {Caron}, \citenamefont {Eckner}, \citenamefont {Hendriks}, \citenamefont {J{\'o}hannesson}, \citenamefont {Ruiz De~Austri},\ and\ \citenamefont {Zaharijas}}]{Caron:2022akb}%
  \BibitemOpen
  \bibfield  {author} {\bibinfo {author} {\bibfnamefont {S.}~\bibnamefont {Caron}}, \bibinfo {author} {\bibfnamefont {C.}~\bibnamefont {Eckner}}, \bibinfo {author} {\bibfnamefont {L.}~\bibnamefont {Hendriks}}, \bibinfo {author} {\bibfnamefont {G.}~\bibnamefont {J{\'o}hannesson}}, \bibinfo {author} {\bibfnamefont {R.}~\bibnamefont {Ruiz De~Austri}},\ and\ \bibinfo {author} {\bibfnamefont {G.}~\bibnamefont {Zaharijas}},\ }\bibfield  {title} {\bibinfo {title} {Mind the gap: The discrepancy between simulation and reality drives interpretations of the {{Galactic Center Excess}}},\ }\href {https://doi.org/10.1088/1475-7516/2023/06/013} {\bibfield  {journal} {\bibinfo  {journal} {Journal of Cosmology and Astroparticle Physics}\ }\textbf {\bibinfo {volume} {2023}}\bibinfo  {number} { (06)},\ \bibinfo {pages} {013}}\BibitemShut {NoStop}%
\bibitem [{\citenamefont {Butter}\ \emph {et~al.}(2023)\citenamefont {Butter}, \citenamefont {Kr{\"a}mer}, \citenamefont {Manconi},\ and\ \citenamefont {Nippel}}]{Butter:2023piw}%
  \BibitemOpen
\bibfield  {number} {  }\bibfield  {author} {\bibinfo {author} {\bibfnamefont {A.}~\bibnamefont {Butter}}, \bibinfo {author} {\bibfnamefont {M.}~\bibnamefont {Kr{\"a}mer}}, \bibinfo {author} {\bibfnamefont {S.}~\bibnamefont {Manconi}},\ and\ \bibinfo {author} {\bibfnamefont {K.}~\bibnamefont {Nippel}},\ }\bibfield  {title} {\bibinfo {title} {Searching for dark matter subhalos in the {{Fermi-LAT}} catalog with {{Bayesian}} neural networks},\ }\href {https://doi.org/10.1088/1475-7516/2023/07/033} {\bibfield  {journal} {\bibinfo  {journal} {Journal of Cosmology and Astroparticle Physics}\ }\textbf {\bibinfo {volume} {2023}},\ \bibinfo {pages} {033}}\BibitemShut {NoStop}%
\bibitem [{\citenamefont {Amerio}\ \emph {et~al.}(2023)\citenamefont {Amerio}, \citenamefont {Cuoco},\ and\ \citenamefont {Fornengo}}]{Amerio:2023uet}%
  \BibitemOpen
  \bibfield  {author} {\bibinfo {author} {\bibfnamefont {A.}~\bibnamefont {Amerio}}, \bibinfo {author} {\bibfnamefont {A.}~\bibnamefont {Cuoco}},\ and\ \bibinfo {author} {\bibfnamefont {N.}~\bibnamefont {Fornengo}},\ }\bibfield  {title} {\bibinfo {title} {Extracting the gamma-ray source-count distribution below the {{Fermi-LAT}} detection limit with deep learning},\ }\href {https://doi.org/10.1088/1475-7516/2023/09/029} {\bibfield  {journal} {\bibinfo  {journal} {Journal of Cosmology and Astroparticle Physics}\ }\textbf {\bibinfo {volume} {2023}},\ \bibinfo {pages} {029}}\BibitemShut {NoStop}%
\bibitem [{\citenamefont {Christy}\ \emph {et~al.}(2024)\citenamefont {Christy}, \citenamefont {Baxter},\ and\ \citenamefont {Kumar}}]{Christy:2024gsl}%
  \BibitemOpen
  \bibfield  {author} {\bibinfo {author} {\bibfnamefont {K.}~\bibnamefont {Christy}}, \bibinfo {author} {\bibfnamefont {E.~J.}\ \bibnamefont {Baxter}},\ and\ \bibinfo {author} {\bibfnamefont {J.}~\bibnamefont {Kumar}},\ }\bibfield  {title} {\bibinfo {title} {Applying simulation-based inference to spectral and spatial information from the {{Galactic Center}} gamma-ray excess},\ }\href {https://doi.org/10.1088/1475-7516/2024/07/066} {\bibfield  {journal} {\bibinfo  {journal} {Journal of Cosmology and Astroparticle Physics}\ }\textbf {\bibinfo {volume} {2024}},\ \bibinfo {pages} {066}}\BibitemShut {NoStop}%
\bibitem [{\citenamefont {Wolf}\ \emph {et~al.}(2024)\citenamefont {Wolf}, \citenamefont {List}, \citenamefont {Rodd},\ and\ \citenamefont {Hahn}}]{Wolf:2024oqb}%
  \BibitemOpen
  \bibfield  {author} {\bibinfo {author} {\bibfnamefont {F.}~\bibnamefont {Wolf}}, \bibinfo {author} {\bibfnamefont {F.}~\bibnamefont {List}}, \bibinfo {author} {\bibfnamefont {N.~L.}\ \bibnamefont {Rodd}},\ and\ \bibinfo {author} {\bibfnamefont {O.}~\bibnamefont {Hahn}},\ }\href {https://doi.org/10.48550/arXiv.2401.03336} {\bibinfo {title} {A deep learning framework for jointly extracting spectra and source-count distributions in astronomy}} (\bibinfo {year} {2024})\BibitemShut {NoStop}%
\bibitem [{\citenamefont {List}\ \emph {et~al.}(2025)\citenamefont {List}, \citenamefont {Park}, \citenamefont {Rodd}, \citenamefont {Schoen},\ and\ \citenamefont {Wolf}}]{List:2025qbx}%
  \BibitemOpen
  \bibfield  {author} {\bibinfo {author} {\bibfnamefont {F.}~\bibnamefont {List}}, \bibinfo {author} {\bibfnamefont {Y.}~\bibnamefont {Park}}, \bibinfo {author} {\bibfnamefont {N.~L.}\ \bibnamefont {Rodd}}, \bibinfo {author} {\bibfnamefont {E.}~\bibnamefont {Schoen}},\ and\ \bibinfo {author} {\bibfnamefont {F.}~\bibnamefont {Wolf}},\ }\href {https://doi.org/10.48550/arXiv.2507.17804} {\bibinfo {title} {On the {{Energy Distribution}} of the {{Galactic Center Excess}}' {{Sources}}}} (\bibinfo {year} {2025}),\ \Eprint {https://arxiv.org/abs/2507.17804} {arXiv:2507.17804 [astro-ph]} \BibitemShut {NoStop}%
\bibitem [{\citenamefont {Gormley}\ \emph {et~al.}(2023)\citenamefont {Gormley}, \citenamefont {Murphy},\ and\ \citenamefont {Raftery}}]{gormleyModelBasedClustering2023}%
  \BibitemOpen
  \bibfield  {author} {\bibinfo {author} {\bibfnamefont {I.~C.}\ \bibnamefont {Gormley}}, \bibinfo {author} {\bibfnamefont {T.~B.}\ \bibnamefont {Murphy}},\ and\ \bibinfo {author} {\bibfnamefont {A.~E.}\ \bibnamefont {Raftery}},\ }\bibfield  {title} {\bibinfo {title} {Model-{{Based Clustering}}},\ }\href {https://doi.org/10.1146/annurev-statistics-033121-115326} {\bibfield  {journal} {\bibinfo  {journal} {Annual Review of Statistics and Its Application}\ }\textbf {\bibinfo {volume} {10}},\ \bibinfo {pages} {573} (\bibinfo {year} {2023})}\BibitemShut {NoStop}%
\bibitem [{\citenamefont {Celeux}\ \emph {et~al.}(2018)\citenamefont {Celeux}, \citenamefont {Kamary}, \citenamefont {{Malsiner-Walli}}, \citenamefont {Marin},\ and\ \citenamefont {Robert}}]{celeuxComputationalSolutionsBayesian2018}%
  \BibitemOpen
  \bibfield  {author} {\bibinfo {author} {\bibfnamefont {G.}~\bibnamefont {Celeux}}, \bibinfo {author} {\bibfnamefont {K.}~\bibnamefont {Kamary}}, \bibinfo {author} {\bibfnamefont {G.}~\bibnamefont {{Malsiner-Walli}}}, \bibinfo {author} {\bibfnamefont {J.-M.}\ \bibnamefont {Marin}},\ and\ \bibinfo {author} {\bibfnamefont {C.~P.}\ \bibnamefont {Robert}},\ }\href {https://doi.org/10.48550/arXiv.1812.07240} {\bibinfo {title} {Computational {{Solutions}} for {{Bayesian Inference}} in {{Mixture Models}}}} (\bibinfo {year} {2018})\BibitemShut {NoStop}%
\bibitem [{\citenamefont {Rousseau}\ and\ \citenamefont {Mengersen}(2011)}]{rousseauAsymptoticBehaviourPosterior2011}%
  \BibitemOpen
  \bibfield  {author} {\bibinfo {author} {\bibfnamefont {J.}~\bibnamefont {Rousseau}}\ and\ \bibinfo {author} {\bibfnamefont {K.}~\bibnamefont {Mengersen}},\ }\bibfield  {title} {\bibinfo {title} {Asymptotic {{Behaviour}} of the {{Posterior Distribution}} in {{Overfitted Mixture Models}}},\ }\href {https://doi.org/10.1111/j.1467-9868.2011.00781.x} {\bibfield  {journal} {\bibinfo  {journal} {Journal of the Royal Statistical Society Series B: Statistical Methodology}\ }\textbf {\bibinfo {volume} {73}},\ \bibinfo {pages} {689} (\bibinfo {year} {2011})}\BibitemShut {NoStop}%
\bibitem [{\citenamefont {{Malsiner-Walli}}\ \emph {et~al.}(2016)\citenamefont {{Malsiner-Walli}}, \citenamefont {{Fr{\"u}hwirth-Schnatter}},\ and\ \citenamefont {Gr{\"u}n}}]{malsiner-walliModelbasedClusteringBased2016}%
  \BibitemOpen
  \bibfield  {author} {\bibinfo {author} {\bibfnamefont {G.}~\bibnamefont {{Malsiner-Walli}}}, \bibinfo {author} {\bibfnamefont {S.}~\bibnamefont {{Fr{\"u}hwirth-Schnatter}}},\ and\ \bibinfo {author} {\bibfnamefont {B.}~\bibnamefont {Gr{\"u}n}},\ }\bibfield  {title} {\bibinfo {title} {Model-based clustering based on sparse finite {{Gaussian}} mixtures},\ }\href {https://doi.org/10.1007/s11222-014-9500-2} {\bibfield  {journal} {\bibinfo  {journal} {Statistics and Computing}\ }\textbf {\bibinfo {volume} {26}},\ \bibinfo {pages} {303} (\bibinfo {year} {2016})}\BibitemShut {NoStop}%
\bibitem [{\citenamefont {Mitchell}\ and\ \citenamefont {Beauchamp}(1988)}]{mitchellBayesianVariableSelection1988}%
  \BibitemOpen
  \bibfield  {author} {\bibinfo {author} {\bibfnamefont {T.~J.}\ \bibnamefont {Mitchell}}\ and\ \bibinfo {author} {\bibfnamefont {J.~J.}\ \bibnamefont {Beauchamp}},\ }\bibfield  {title} {\bibinfo {title} {Bayesian {{Variable Selection}} in {{Linear Regression}}},\ }\href {https://doi.org/10.1080/01621459.1988.10478694} {\bibfield  {journal} {\bibinfo  {journal} {Journal of the American Statistical Association}\ }\textbf {\bibinfo {volume} {83}},\ \bibinfo {pages} {1023} (\bibinfo {year} {1988})}\BibitemShut {NoStop}%
\bibitem [{\citenamefont {Mak}\ \emph {et~al.}(2021)\citenamefont {Mak}, \citenamefont {Zaiser},\ and\ \citenamefont {Ong}}]{pmlr-v139-mak21a}%
  \BibitemOpen
  \bibfield  {author} {\bibinfo {author} {\bibfnamefont {C.}~\bibnamefont {Mak}}, \bibinfo {author} {\bibfnamefont {F.}~\bibnamefont {Zaiser}},\ and\ \bibinfo {author} {\bibfnamefont {L.}~\bibnamefont {Ong}},\ }\bibfield  {title} {\bibinfo {title} {Nonparametric hamiltonian monte carlo},\ }in\ \href@noop {} {\emph {\bibinfo {booktitle} {Proceedings of the 38th International Conference on Machine Learning}}},\ \bibinfo {series} {Proceedings of Machine Learning Research}, Vol.\ \bibinfo {volume} {139},\ \bibinfo {editor} {edited by\ \bibinfo {editor} {\bibfnamefont {M.}~\bibnamefont {Meila}}\ and\ \bibinfo {editor} {\bibfnamefont {T.}~\bibnamefont {Zhang}}}\ (\bibinfo  {publisher} {PMLR},\ \bibinfo {year} {2021})\ pp.\ \bibinfo {pages} {7336--7347}\BibitemShut {NoStop}%
\bibitem [{\citenamefont {Robnik}\ \emph {et~al.}(2024)\citenamefont {Robnik}, \citenamefont {De~Luca}, \citenamefont {Silverstein},\ and\ \citenamefont {Seljak}}]{10.5555/3648699.3649010}%
  \BibitemOpen
  \bibfield  {author} {\bibinfo {author} {\bibfnamefont {J.}~\bibnamefont {Robnik}}, \bibinfo {author} {\bibfnamefont {G.~B.}\ \bibnamefont {De~Luca}}, \bibinfo {author} {\bibfnamefont {E.}~\bibnamefont {Silverstein}},\ and\ \bibinfo {author} {\bibfnamefont {U.}~\bibnamefont {Seljak}},\ }\bibfield  {title} {\bibinfo {title} {Microcanonical hamiltonian monte carlo},\ }\href@noop {} {\bibfield  {journal} {\bibinfo  {journal} {Journal of Machine Learning Research}\ }\textbf {\bibinfo {volume} {24}} (\bibinfo {year} {2024})}\BibitemShut {NoStop}%
\bibitem [{\citenamefont {Robnik}\ and\ \citenamefont {Seljak}(2024)}]{robnik2024fluctuation}%
  \BibitemOpen
  \bibfield  {author} {\bibinfo {author} {\bibfnamefont {J.}~\bibnamefont {Robnik}}\ and\ \bibinfo {author} {\bibfnamefont {U.}~\bibnamefont {Seljak}},\ }\bibfield  {title} {\bibinfo {title} {Fluctuation without dissipation: {{Microcanonical}} langevin monte carlo},\ }in\ \href@noop {} {\emph {\bibinfo {booktitle} {Symposium on Advances in Approximate Bayesian Inference}}}\ (\bibinfo  {publisher} {PMLR},\ \bibinfo {year} {2024})\ pp.\ \bibinfo {pages} {111--126}\BibitemShut {NoStop}%
\bibitem [{\citenamefont {Bayer}\ \emph {et~al.}(2023)\citenamefont {Bayer}, \citenamefont {Seljak},\ and\ \citenamefont {Modi}}]{Bayer:2023rmj}%
  \BibitemOpen
  \bibfield  {author} {\bibinfo {author} {\bibfnamefont {A.~E.}\ \bibnamefont {Bayer}}, \bibinfo {author} {\bibfnamefont {U.}~\bibnamefont {Seljak}},\ and\ \bibinfo {author} {\bibfnamefont {C.}~\bibnamefont {Modi}},\ }\href {https://doi.org/10.48550/arXiv.2307.09504} {\bibinfo {title} {Field-{{Level Inference}} with {{Microcanonical Langevin Monte Carlo}}}} (\bibinfo {year} {2023})\BibitemShut {NoStop}%
\bibitem [{\citenamefont {Cabezas}\ \emph {et~al.}(2024)\citenamefont {Cabezas}, \citenamefont {Corenflos}, \citenamefont {Lao}, \citenamefont {Louf}, \citenamefont {Carnec}, \citenamefont {Chaudhari}, \citenamefont {{Cohn-Gordon}}, \citenamefont {Coullon}, \citenamefont {Deng}, \citenamefont {Duffield}, \citenamefont {{Dur{\'a}n-Mart{\'i}n}}, \citenamefont {Elantkowski}, \citenamefont {{Foreman-Mackey}}, \citenamefont {Gregori}, \citenamefont {Iguaran}, \citenamefont {Kumar}, \citenamefont {Lysy}, \citenamefont {Murphy}, \citenamefont {Orduz}, \citenamefont {Patel}, \citenamefont {Wang},\ and\ \citenamefont {Zinkov}}]{cabezasBlackJAXComposableBayesian2024}%
  \BibitemOpen
  \bibfield  {author} {\bibinfo {author} {\bibfnamefont {A.}~\bibnamefont {Cabezas}}, \bibinfo {author} {\bibfnamefont {A.}~\bibnamefont {Corenflos}}, \bibinfo {author} {\bibfnamefont {J.}~\bibnamefont {Lao}}, \bibinfo {author} {\bibfnamefont {R.}~\bibnamefont {Louf}}, \bibinfo {author} {\bibfnamefont {A.}~\bibnamefont {Carnec}}, \bibinfo {author} {\bibfnamefont {K.}~\bibnamefont {Chaudhari}}, \bibinfo {author} {\bibfnamefont {R.}~\bibnamefont {{Cohn-Gordon}}}, \bibinfo {author} {\bibfnamefont {J.}~\bibnamefont {Coullon}}, \bibinfo {author} {\bibfnamefont {W.}~\bibnamefont {Deng}}, \bibinfo {author} {\bibfnamefont {S.}~\bibnamefont {Duffield}}, \bibinfo {author} {\bibfnamefont {G.}~\bibnamefont {{Dur{\'a}n-Mart{\'i}n}}}, \bibinfo {author} {\bibfnamefont {M.}~\bibnamefont {Elantkowski}}, \bibinfo {author} {\bibfnamefont {D.}~\bibnamefont {{Foreman-Mackey}}}, \bibinfo {author} {\bibfnamefont {M.}~\bibnamefont {Gregori}}, \bibinfo {author} {\bibfnamefont {C.}~\bibnamefont {Iguaran}}, \bibinfo {author}
  {\bibfnamefont {R.}~\bibnamefont {Kumar}}, \bibinfo {author} {\bibfnamefont {M.}~\bibnamefont {Lysy}}, \bibinfo {author} {\bibfnamefont {K.}~\bibnamefont {Murphy}}, \bibinfo {author} {\bibfnamefont {J.~C.}\ \bibnamefont {Orduz}}, \bibinfo {author} {\bibfnamefont {K.}~\bibnamefont {Patel}}, \bibinfo {author} {\bibfnamefont {X.}~\bibnamefont {Wang}},\ and\ \bibinfo {author} {\bibfnamefont {R.}~\bibnamefont {Zinkov}},\ }\href {https://doi.org/10.48550/arXiv.2402.10797} {\bibinfo {title} {{{BlackJAX}}: {{Composable Bayesian}} inference in {{JAX}}}} (\bibinfo {year} {2024})\BibitemShut {NoStop}%
\bibitem [{\citenamefont {Wik}\ \emph {et~al.}(2014)\citenamefont {Wik}, \citenamefont {Hornstrup}, \citenamefont {Molendi}, \citenamefont {Madejski}, \citenamefont {Harrison}, \citenamefont {Zoglauer}, \citenamefont {Grefenstette}, \citenamefont {Gastaldello}, \citenamefont {Madsen}, \citenamefont {Westergaard}, \citenamefont {Ferreira}, \citenamefont {Kitaguchi}, \citenamefont {Pedersen}, \citenamefont {Boggs}, \citenamefont {Christensen}, \citenamefont {Craig}, \citenamefont {Hailey}, \citenamefont {Stern},\ and\ \citenamefont {Zhang}}]{Wik:2014boa}%
  \BibitemOpen
  \bibfield  {author} {\bibinfo {author} {\bibfnamefont {D.~R.}\ \bibnamefont {Wik}}, \bibinfo {author} {\bibfnamefont {A.}~\bibnamefont {Hornstrup}}, \bibinfo {author} {\bibfnamefont {S.}~\bibnamefont {Molendi}}, \bibinfo {author} {\bibfnamefont {G.}~\bibnamefont {Madejski}}, \bibinfo {author} {\bibfnamefont {F.~A.}\ \bibnamefont {Harrison}}, \bibinfo {author} {\bibfnamefont {A.}~\bibnamefont {Zoglauer}}, \bibinfo {author} {\bibfnamefont {B.~W.}\ \bibnamefont {Grefenstette}}, \bibinfo {author} {\bibfnamefont {F.}~\bibnamefont {Gastaldello}}, \bibinfo {author} {\bibfnamefont {K.~K.}\ \bibnamefont {Madsen}}, \bibinfo {author} {\bibfnamefont {N.~J.}\ \bibnamefont {Westergaard}}, \bibinfo {author} {\bibfnamefont {D.~D.~M.}\ \bibnamefont {Ferreira}}, \bibinfo {author} {\bibfnamefont {T.}~\bibnamefont {Kitaguchi}}, \bibinfo {author} {\bibfnamefont {K.}~\bibnamefont {Pedersen}}, \bibinfo {author} {\bibfnamefont {S.~E.}\ \bibnamefont {Boggs}}, \bibinfo {author} {\bibfnamefont {F.~E.}\ \bibnamefont {Christensen}},
  \bibinfo {author} {\bibfnamefont {W.~W.}\ \bibnamefont {Craig}}, \bibinfo {author} {\bibfnamefont {C.~J.}\ \bibnamefont {Hailey}}, \bibinfo {author} {\bibfnamefont {D.}~\bibnamefont {Stern}},\ and\ \bibinfo {author} {\bibfnamefont {W.~W.}\ \bibnamefont {Zhang}},\ }\bibfield  {title} {\bibinfo {title} {{{NuSTAR Observations}} of the {{Bullet Cluster}}: {{Constraints}} on {{Inverse Compton Emission}}},\ }\href {https://doi.org/10.1088/0004-637X/792/1/48} {\bibfield  {journal} {\bibinfo  {journal} {The Astrophysical Journal}\ }\textbf {\bibinfo {volume} {792}},\ \bibinfo {pages} {48} (\bibinfo {year} {2014})}\BibitemShut {NoStop}%
\bibitem [{\citenamefont {Madsen}\ \emph {et~al.}(2015)\citenamefont {Madsen}, \citenamefont {Harrison}, \citenamefont {Markwardt}, \citenamefont {An}, \citenamefont {Grefenstette}, \citenamefont {Bachetti}, \citenamefont {Miyasaka}, \citenamefont {Kitaguchi}, \citenamefont {Bhalerao}, \citenamefont {Boggs}, \citenamefont {Christensen}, \citenamefont {Craig}, \citenamefont {Forster}, \citenamefont {Fuerst}, \citenamefont {Hailey}, \citenamefont {Perri}, \citenamefont {Puccetti}, \citenamefont {Rana}, \citenamefont {Stern}, \citenamefont {Walton}, \citenamefont {J{\o}rgen~Westergaard},\ and\ \citenamefont {Zhang}}]{Madsen:2015jea}%
  \BibitemOpen
  \bibfield  {author} {\bibinfo {author} {\bibfnamefont {K.~K.}\ \bibnamefont {Madsen}}, \bibinfo {author} {\bibfnamefont {F.~A.}\ \bibnamefont {Harrison}}, \bibinfo {author} {\bibfnamefont {C.~B.}\ \bibnamefont {Markwardt}}, \bibinfo {author} {\bibfnamefont {H.}~\bibnamefont {An}}, \bibinfo {author} {\bibfnamefont {B.~W.}\ \bibnamefont {Grefenstette}}, \bibinfo {author} {\bibfnamefont {M.}~\bibnamefont {Bachetti}}, \bibinfo {author} {\bibfnamefont {H.}~\bibnamefont {Miyasaka}}, \bibinfo {author} {\bibfnamefont {T.}~\bibnamefont {Kitaguchi}}, \bibinfo {author} {\bibfnamefont {V.}~\bibnamefont {Bhalerao}}, \bibinfo {author} {\bibfnamefont {S.}~\bibnamefont {Boggs}}, \bibinfo {author} {\bibfnamefont {F.~E.}\ \bibnamefont {Christensen}}, \bibinfo {author} {\bibfnamefont {W.~W.}\ \bibnamefont {Craig}}, \bibinfo {author} {\bibfnamefont {K.}~\bibnamefont {Forster}}, \bibinfo {author} {\bibfnamefont {F.}~\bibnamefont {Fuerst}}, \bibinfo {author} {\bibfnamefont {C.~J.}\ \bibnamefont {Hailey}}, \bibinfo {author}
  {\bibfnamefont {M.}~\bibnamefont {Perri}}, \bibinfo {author} {\bibfnamefont {S.}~\bibnamefont {Puccetti}}, \bibinfo {author} {\bibfnamefont {V.}~\bibnamefont {Rana}}, \bibinfo {author} {\bibfnamefont {D.}~\bibnamefont {Stern}}, \bibinfo {author} {\bibfnamefont {D.~J.}\ \bibnamefont {Walton}}, \bibinfo {author} {\bibfnamefont {N.}~\bibnamefont {J{\o}rgen~Westergaard}},\ and\ \bibinfo {author} {\bibfnamefont {W.~W.}\ \bibnamefont {Zhang}},\ }\bibfield  {title} {\bibinfo {title} {Calibration of the {{NuSTAR High-energy Focusing X-ray Telescope}}.},\ }\href {https://doi.org/10.1088/0067-0049/220/1/8} {\bibfield  {journal} {\bibinfo  {journal} {The Astrophysical Journal Supplement Series}\ }\textbf {\bibinfo {volume} {220}},\ \bibinfo {pages} {8} (\bibinfo {year} {2015})}\BibitemShut {NoStop}%
\bibitem [{\citenamefont {Madsen}\ \emph {et~al.}(2017)\citenamefont {Madsen}, \citenamefont {Christensen}, \citenamefont {Craig}, \citenamefont {Forster}, \citenamefont {Grefenstette}, \citenamefont {Harrison}, \citenamefont {Miyasaka},\ and\ \citenamefont {Rana}}]{madsenObservationalArtifactsNuSTAR2017}%
  \BibitemOpen
  \bibfield  {author} {\bibinfo {author} {\bibfnamefont {K.~K.}\ \bibnamefont {Madsen}}, \bibinfo {author} {\bibfnamefont {F.~E.}\ \bibnamefont {Christensen}}, \bibinfo {author} {\bibfnamefont {W.~W.}\ \bibnamefont {Craig}}, \bibinfo {author} {\bibfnamefont {K.~W.}\ \bibnamefont {Forster}}, \bibinfo {author} {\bibfnamefont {B.~W.}\ \bibnamefont {Grefenstette}}, \bibinfo {author} {\bibfnamefont {F.~A.}\ \bibnamefont {Harrison}}, \bibinfo {author} {\bibfnamefont {H.}~\bibnamefont {Miyasaka}},\ and\ \bibinfo {author} {\bibfnamefont {V.}~\bibnamefont {Rana}},\ }\href {https://doi.org/10.48550/arXiv.1711.02719} {\bibinfo {title} {Observational {{Artifacts}} of {{NuSTAR}}: {{Ghost Rays}} and {{Stray Light}}}} (\bibinfo {year} {2017})\BibitemShut {NoStop}%
\bibitem [{\citenamefont {Harrison}\ \emph {et~al.}(2013)\citenamefont {Harrison}, \citenamefont {Craig}, \citenamefont {Christensen}, \citenamefont {Hailey}, \citenamefont {Zhang}, \citenamefont {Boggs}, \citenamefont {Stern}, \citenamefont {Cook}, \citenamefont {Forster}, \citenamefont {Giommi}, \citenamefont {Grefenstette}, \citenamefont {Kim}, \citenamefont {Kitaguchi}, \citenamefont {Koglin}, \citenamefont {Madsen}, \citenamefont {Mao}, \citenamefont {Miyasaka}, \citenamefont {Mori}, \citenamefont {Perri}, \citenamefont {Pivovaroff}, \citenamefont {Puccetti}, \citenamefont {Rana}, \citenamefont {Westergaard}, \citenamefont {Willis}, \citenamefont {Zoglauer}, \citenamefont {An}, \citenamefont {Bachetti}, \citenamefont {Barri{\`e}re}, \citenamefont {Bellm}, \citenamefont {Bhalerao}, \citenamefont {Brejnholt}, \citenamefont {Fuerst}, \citenamefont {Liebe}, \citenamefont {Markwardt}, \citenamefont {Nynka}, \citenamefont {Vogel}, \citenamefont {Walton}, \citenamefont {Wik}, \citenamefont {Alexander},
  \citenamefont {Cominsky}, \citenamefont {Hornschemeier}, \citenamefont {Hornstrup}, \citenamefont {Kaspi}, \citenamefont {Madejski}, \citenamefont {Matt}, \citenamefont {Molendi}, \citenamefont {Smith}, \citenamefont {Tomsick}, \citenamefont {Ajello}, \citenamefont {Ballantyne}, \citenamefont {Balokovi{\'c}}, \citenamefont {Barret}, \citenamefont {Bauer}, \citenamefont {Blandford}, \citenamefont {Brandt}, \citenamefont {Brenneman}, \citenamefont {Chiang}, \citenamefont {Chakrabarty}, \citenamefont {Chenevez}, \citenamefont {Comastri}, \citenamefont {Dufour}, \citenamefont {Elvis}, \citenamefont {Fabian}, \citenamefont {Farrah}, \citenamefont {Fryer}, \citenamefont {Gotthelf}, \citenamefont {Grindlay}, \citenamefont {Helfand}, \citenamefont {Krivonos}, \citenamefont {Meier}, \citenamefont {Miller}, \citenamefont {Natalucci}, \citenamefont {Ogle}, \citenamefont {Ofek}, \citenamefont {Ptak}, \citenamefont {Reynolds}, \citenamefont {Rigby}, \citenamefont {Tagliaferri}, \citenamefont {Thorsett}, \citenamefont
  {Treister},\ and\ \citenamefont {Urry}}]{NuSTAR:2013yza}%
  \BibitemOpen
  \bibfield  {author} {\bibinfo {author} {\bibfnamefont {F.~A.}\ \bibnamefont {Harrison}}, \bibinfo {author} {\bibfnamefont {W.~W.}\ \bibnamefont {Craig}}, \bibinfo {author} {\bibfnamefont {F.~E.}\ \bibnamefont {Christensen}}, \bibinfo {author} {\bibfnamefont {C.~J.}\ \bibnamefont {Hailey}}, \bibinfo {author} {\bibfnamefont {W.~W.}\ \bibnamefont {Zhang}}, \bibinfo {author} {\bibfnamefont {S.~E.}\ \bibnamefont {Boggs}}, \bibinfo {author} {\bibfnamefont {D.}~\bibnamefont {Stern}}, \bibinfo {author} {\bibfnamefont {W.~R.}\ \bibnamefont {Cook}}, \bibinfo {author} {\bibfnamefont {K.}~\bibnamefont {Forster}}, \bibinfo {author} {\bibfnamefont {P.}~\bibnamefont {Giommi}}, \bibinfo {author} {\bibfnamefont {B.~W.}\ \bibnamefont {Grefenstette}}, \bibinfo {author} {\bibfnamefont {Y.}~\bibnamefont {Kim}}, \bibinfo {author} {\bibfnamefont {T.}~\bibnamefont {Kitaguchi}}, \bibinfo {author} {\bibfnamefont {J.~E.}\ \bibnamefont {Koglin}}, \bibinfo {author} {\bibfnamefont {K.~K.}\ \bibnamefont {Madsen}}, \bibinfo {author}
  {\bibfnamefont {P.~H.}\ \bibnamefont {Mao}}, \bibinfo {author} {\bibfnamefont {H.}~\bibnamefont {Miyasaka}}, \bibinfo {author} {\bibfnamefont {K.}~\bibnamefont {Mori}}, \bibinfo {author} {\bibfnamefont {M.}~\bibnamefont {Perri}}, \bibinfo {author} {\bibfnamefont {M.~J.}\ \bibnamefont {Pivovaroff}}, \bibinfo {author} {\bibfnamefont {S.}~\bibnamefont {Puccetti}}, \bibinfo {author} {\bibfnamefont {V.~R.}\ \bibnamefont {Rana}}, \bibinfo {author} {\bibfnamefont {N.~J.}\ \bibnamefont {Westergaard}}, \bibinfo {author} {\bibfnamefont {J.}~\bibnamefont {Willis}}, \bibinfo {author} {\bibfnamefont {A.}~\bibnamefont {Zoglauer}}, \bibinfo {author} {\bibfnamefont {H.}~\bibnamefont {An}}, \bibinfo {author} {\bibfnamefont {M.}~\bibnamefont {Bachetti}}, \bibinfo {author} {\bibfnamefont {N.~M.}\ \bibnamefont {Barri{\`e}re}}, \bibinfo {author} {\bibfnamefont {E.~C.}\ \bibnamefont {Bellm}}, \bibinfo {author} {\bibfnamefont {V.}~\bibnamefont {Bhalerao}}, \bibinfo {author} {\bibfnamefont {N.~F.}\ \bibnamefont {Brejnholt}},
  \bibinfo {author} {\bibfnamefont {F.}~\bibnamefont {Fuerst}}, \bibinfo {author} {\bibfnamefont {C.~C.}\ \bibnamefont {Liebe}}, \bibinfo {author} {\bibfnamefont {C.~B.}\ \bibnamefont {Markwardt}}, \bibinfo {author} {\bibfnamefont {M.}~\bibnamefont {Nynka}}, \bibinfo {author} {\bibfnamefont {J.~K.}\ \bibnamefont {Vogel}}, \bibinfo {author} {\bibfnamefont {D.~J.}\ \bibnamefont {Walton}}, \bibinfo {author} {\bibfnamefont {D.~R.}\ \bibnamefont {Wik}}, \bibinfo {author} {\bibfnamefont {D.~M.}\ \bibnamefont {Alexander}}, \bibinfo {author} {\bibfnamefont {L.~R.}\ \bibnamefont {Cominsky}}, \bibinfo {author} {\bibfnamefont {A.~E.}\ \bibnamefont {Hornschemeier}}, \bibinfo {author} {\bibfnamefont {A.}~\bibnamefont {Hornstrup}}, \bibinfo {author} {\bibfnamefont {V.~M.}\ \bibnamefont {Kaspi}}, \bibinfo {author} {\bibfnamefont {G.~M.}\ \bibnamefont {Madejski}}, \bibinfo {author} {\bibfnamefont {G.}~\bibnamefont {Matt}}, \bibinfo {author} {\bibfnamefont {S.}~\bibnamefont {Molendi}}, \bibinfo {author} {\bibfnamefont
  {D.~M.}\ \bibnamefont {Smith}}, \bibinfo {author} {\bibfnamefont {J.~A.}\ \bibnamefont {Tomsick}}, \bibinfo {author} {\bibfnamefont {M.}~\bibnamefont {Ajello}}, \bibinfo {author} {\bibfnamefont {D.~R.}\ \bibnamefont {Ballantyne}}, \bibinfo {author} {\bibfnamefont {M.}~\bibnamefont {Balokovi{\'c}}}, \bibinfo {author} {\bibfnamefont {D.}~\bibnamefont {Barret}}, \bibinfo {author} {\bibfnamefont {F.~E.}\ \bibnamefont {Bauer}}, \bibinfo {author} {\bibfnamefont {R.~D.}\ \bibnamefont {Blandford}}, \bibinfo {author} {\bibfnamefont {W.~N.}\ \bibnamefont {Brandt}}, \bibinfo {author} {\bibfnamefont {L.~W.}\ \bibnamefont {Brenneman}}, \bibinfo {author} {\bibfnamefont {J.}~\bibnamefont {Chiang}}, \bibinfo {author} {\bibfnamefont {D.}~\bibnamefont {Chakrabarty}}, \bibinfo {author} {\bibfnamefont {J.}~\bibnamefont {Chenevez}}, \bibinfo {author} {\bibfnamefont {A.}~\bibnamefont {Comastri}}, \bibinfo {author} {\bibfnamefont {F.}~\bibnamefont {Dufour}}, \bibinfo {author} {\bibfnamefont {M.}~\bibnamefont {Elvis}}, \bibinfo
  {author} {\bibfnamefont {A.~C.}\ \bibnamefont {Fabian}}, \bibinfo {author} {\bibfnamefont {D.}~\bibnamefont {Farrah}}, \bibinfo {author} {\bibfnamefont {C.~L.}\ \bibnamefont {Fryer}}, \bibinfo {author} {\bibfnamefont {E.~V.}\ \bibnamefont {Gotthelf}}, \bibinfo {author} {\bibfnamefont {J.~E.}\ \bibnamefont {Grindlay}}, \bibinfo {author} {\bibfnamefont {D.~J.}\ \bibnamefont {Helfand}}, \bibinfo {author} {\bibfnamefont {R.}~\bibnamefont {Krivonos}}, \bibinfo {author} {\bibfnamefont {D.~L.}\ \bibnamefont {Meier}}, \bibinfo {author} {\bibfnamefont {J.~M.}\ \bibnamefont {Miller}}, \bibinfo {author} {\bibfnamefont {L.}~\bibnamefont {Natalucci}}, \bibinfo {author} {\bibfnamefont {P.}~\bibnamefont {Ogle}}, \bibinfo {author} {\bibfnamefont {E.~O.}\ \bibnamefont {Ofek}}, \bibinfo {author} {\bibfnamefont {A.}~\bibnamefont {Ptak}}, \bibinfo {author} {\bibfnamefont {S.~P.}\ \bibnamefont {Reynolds}}, \bibinfo {author} {\bibfnamefont {J.~R.}\ \bibnamefont {Rigby}}, \bibinfo {author} {\bibfnamefont {G.}~\bibnamefont
  {Tagliaferri}}, \bibinfo {author} {\bibfnamefont {S.~E.}\ \bibnamefont {Thorsett}}, \bibinfo {author} {\bibfnamefont {E.}~\bibnamefont {Treister}},\ and\ \bibinfo {author} {\bibfnamefont {C.~M.}\ \bibnamefont {Urry}},\ }\bibfield  {title} {\bibinfo {title} {The {{Nuclear Spectroscopic Telescope Array}} ({{NuSTAR}}) {{High-energy X-Ray Mission}}},\ }\href {https://doi.org/10.1088/0004-637X/770/2/103} {\bibfield  {journal} {\bibinfo  {journal} {The Astrophysical Journal}\ }\textbf {\bibinfo {volume} {770}},\ \bibinfo {pages} {103} (\bibinfo {year} {2013})}\BibitemShut {NoStop}%
\end{thebibliography}%

\end{document}